\documentclass[fleqn,usenatbib]{mnras}

\usepackage{newtxtext,newtxmath}

\usepackage[T1]{fontenc}
\usepackage{ae,aecompl}

\usepackage{graphicx}	
\usepackage{amsmath}	
\usepackage{amssymb}	

\title[The local main sequence and its bending]{The main sequence of star forming galaxies I. The local relation and its bending}

\author[P.Popesso et al.]{
P. Popesso,$^{1}$\thanks{E-mail: paola.popesso@tum.de}
A. Concas,$^{1}$
L. Morselli,$^{1}$
C. Schreiber,$^{2}$
G. Rodighiero,$^{3}$
\newauthor
G. Cresci,$^{4}$
S. Belli,$^{5}$
G. Erfanianfar,$^{5}$
C. Mancini,$^{2}$
H. Inami,$^{6}$
M. Dickinson,$^{7}$
\newauthor
O. Ilbert,$^{8}$
M. Pannella,$^{9}$
D. Elbaz,$^{10}$
\\
$^{1}$Excellence cluster Universe, Boltzmannstrasse 2, 85748, Garching bey M\:unchen Germany\\
$^{2}$Leiden Observatory, Leiden University PO Box 9500 2300 RA Leiden\\
$^{3}$Universita degli studi di Padova, vicolo dell'Osservatorio, Padova, Italy\\
$^{4}$ Osservatorio Astronomico di Arcetri, INAF, Largo Enrico Fermi 5, 50125 Firenze FI, Italy\\
$^{5}$ Max Planck f\"ur extraterrestrische Physik, Giessenbachstrasse 1, 85478, Garching bey M\'unchen, Germany\\
$^{6}$ Univ. Lyon, Univ. Lyon1, ENS de Lyon, CNRS, Centre de Recherche Astrophysique de Lyon UMR5574, 69230 Saint-Genis-Laval, France\\
$^{7}$ NOAO 950 North Cherry Ave. Tucson, AZ 85719, USA \\
$^{8}$ Laboratoire d'Astrophysique de Marseille, 38 rue Frederic Joliot Curie, 13388 Marseille \\
$^{9}$ UFaculty of Physics, Ludwig-Maximilians-Universität, Scheinerstr. 1, D-81679 Munich, Germany\\
$^{10}$ Service d'Astrophysique du CEA, 91190 Gif-sur-Yvette, France
}

\date{Accepted XXX. Received YYY; in original form ZZZ}

\pubyear{2018}

\begin{document}
\label{firstpage}
\pagerange{\pageref{firstpage}--\pageref{lastpage}}
\maketitle

\begin{abstract}
By using a set of different SFR indicators, including WISE mid-infrared and H$\alpha$ emission, we study the slope of the Main Sequence (MS) of local star forming galaxies at stellar masses larger than $10^{10} M_{\odot}$. The slope of the relation strongly depends on the SFR indicator used. In all cases, the local MS shows a bending at high stellar masses with respect to the slope obtained in the low mass regime. While the distribution of galaxies in the upper envelope of the MS is consistent with a log-normal distribution, the lower envelope shows an excess of galaxies, which increases as a function of the stellar mass but varies as a function of the SFR indicator used. The scatter of the best log-normal distribution increases with stellar mass from $\sim 0.3$ dex at $10^{10} M_{\odot}$ to $\sim 0.45$ at $10^{11} M_{\odot}$. The MS high-mass end is dominated by central galaxies of group sized halos with a red bulge and a disk redder than the lower mass counterparts. We argue that the MS bending in this region is due to two processes: {\it{i)}} the formation of a bulge component as a consequence of the increased merger activity in groups, and {\it{ii)}} the cold gas starvation induced by the hot halo environment, which cuts off the gas inflow onto the disk. Similarly, the increase of the MS scatter at high stellar masses would be explained by the larger spread of star formation histories of central group and cluster galaxies with respect to lower mass systems.
\end{abstract}

\begin{keywords}
galaxies: evolution -- galaxies: star formation -- galaxies: starburst -- galaxies: groups -- galaxies: haloes
\end{keywords}



\section{Introduction}

The formation and assembly of the stellar content of galaxies remain at the heart of galaxy evolution studies. Recent advances have led to an emerging picture where most galaxies form stars at a level dictated mainly by their stellar masses, and regulated by secular processes. This is seen as a rather tight relation between galaxy star formation rate (SFR) and stellar mass, so called main sequence of star forming galaxies, in place from redshift $\sim$ 0 up to $\sim$ 4 (MS; e.g.,  Brinchmann et al. 2004;  Noeske et al. 2007; Elbaz et al. 2007; Daddi et al. 2007; Pannella et al. 2009; Magdis et al. 2010; Gonzalez et al. 2010, Schreiber et al. 2015). The SFR increases with the stellar mass ($M*$) as a power law, $SFR \propto M*^{\alpha}$, with an intrinsic scatter of about 0.2-0.3 dex for moderate to relatively low stellar mass galaxies (Whitaker et al. 2015; Speagle et al. 2014). Measurements of the slope $\alpha$ vary widely in the literature, ranging between $0.6–-1.2$ (see summary in Speagle et al. 2014). The observed relation suggests that prior to the shutdown of star formation, galaxy star formation histories are predominantly regular and smoothly declining on mass-dependent timescales (see also Heavens et al 2004), rather than driven by stochastic events like major mergers and starbursts. However, Ilbert et al. (2015) suggest that the scatter of the relation is not constant but it increases as a function of stellar mass, at least up to $z\sim$ 1.4, reflecting a larger variety of star formation histories for the most massive galaxies.  

Several studies suggest also that the relation is not a power law but it exhibits a curvature with a flatter slope at the high-mass end with respect to the low mass regime (Whitaker et al. 2015, Schreiber et al. 2015, Lee et al. 2015, Tomczak et al. 2016) though bending is not found in other studies (e.g., Speagle et al. 2014; Rodighiero et al. 2014). 

Recently Renzini \& Peng (2015, hereafter RP15) propose two different definitions of the MS: the ridge line connecting the peak of the 3D number density distribution of galaxies over the log(SFR)-log(M*) plane, or of the similar 3D distribution where the $z$ coordinate is given by the product of the number of galaxies  times their SFR. Such definitions are defined as "objective" because they do not require any selection of the SF galaxy population and do not make any assumption on the shape of SFR distribution around the MS. The two definitions give nearly parallel relations in the SDSS galaxy spectroscopic sample at $z < 0.085$ with slope $0.76\pm0.06$ over the stellar mass range $10^{8}-10^{10.5} M_{\odot}$, without any particular bending as in Whitaker et al. (2014, see also Magnelli et al. 2014). 

The MS relations retrieved in the literature differ for a variety of reasons, including how galaxies are selected in the first place. For example, pre-selecting MS galaxies, by separating {\it quenched} from star forming systems, may include galaxies on their way to be quenched with low but still detectable SFR (for example using a color-selection, such as the  $UVJ$ selection). Such inclusion would have the effect of flattening the SFR$-M*$ relation. 

An additional source of discrepancy might be due to the method used to locate the MS. The SFR distribution in the MS region is found in most of the cases to be log-normal (e.g. RP15). However, for such distribution, mean, median and mode are different and would, then, provide different MS locations. Thus, comparing MS defined as the median of the log-normal distribution (the peak in the log(SFR)-log(mass) plane) might differ from the results retrieved in the stacking analysis and based on the mean galaxy SFR.
In addition, if the SFR distribution around the MS deviates from a log-normal distribution, the discrepancy of the different indicators might further diverge. Brinchmann et al. (2004) show clearly that in the local Universe the lower envelope of the MS in the SDSS spectroscopic sample, in particular, at the high mass end, exhibits a valley between the MS and the quiescence region, analogously to the so called "green valley" in the mass-color diagram. These excesses might affect, to different extend, the indicators used to identify the relation.

It is adamant, then, that knowing the SFR distribution in the MS region is the most robust way to properly study the MS relation and its scatter. To this aim, in this paper we present the analysis of the SFR distribution in the MS region for galaxies with masses above $10^{10} M_{\odot}$ in the local Universe. Such distribution is used to define the MS location and its scatter and to investigate the biases due to the SFR indicators and the method used to locate the mean. To this aim we use the most robust and reliable SFR indicators available in the local Universe, either the dust corrected H${\alpha}$-based SFR, or the combination of the UV light emitted by young stars and the IR luminosity, accounting for the UV component absorbed and reprocessed by dust. In particular, in the local Universe we use two different samples: the SDSS spectroscopic sample with SFR based mainly on corrected H$\alpha$ SFR (Brinchmann et al. 2004) and the WISE sample matched to the SDSS sample of S16, with SFR based on WISE 22 $\mu$m data, when available, and dust corrected UV-based SFR.  To further check our results we use also the far infrared H-ATLAS {\it{Herschel}}/SPIRE-based SFR  of Valiante et al. (2016).  

The paper is structured as follows. Section 1 describes our dataset. Section 2 presents our method. Section 3 shows our results and Section 4 contains the summary of the findings and the discussion. We  assume  a  $\Lambda$CDM  cosmology  with  $\Omega_M=0.3$, $\Omega_{\Lambda}=0.7$ and H$_0=70$ $km/s/Mpc$ throughout the paper.

\section{Data and sample selection}

In the following section we describe the local galaxy sample, which is defined on the basis of different catalogs and SFR indicators. The comparison of the different SFR indicators is shown in the Appendix. In this section we describe the main results of such comparison. 

\subsection{The local galaxy sample}

\subsubsection{The SDSS dataset}
The first local galaxy sample is drawn from the Sloan Digital Sky Survey, (SDSS, York et al. 2000). In particular, we use the spectroscopic catalog containing $\sim 930,000$ spectra belonging to the seventh data release (DR7, Abazajian et al. 2009). The spectra cover a wavelength range  from $3800$ to $9200$ \AA. They are obtained with $3''$ diameter aperture fibers. Further details concerning the DR7 spectra can be found at http://www.sdss.org/dr7/. The choice of the DR7 sample is dictated by the use of the  SFR and $M_{\star}$ measurements taken from the MPA-JHU catalog\footnote{http://www.mpa-garching.mpg.de/SDSS/DR7/}, based on DR7. 
The stellar masses are obtained from a fit to the spectral energy distribution (SED) by using the SDSS broad-band optical photometry (see Kauffmann et al. 2003a and Salim et al. 2005 for details). The SFR measurements are based on the Brinchmann et al. (2004) approach. The H$\alpha$ emission line luminosity is used to determine the SFRs for the star forming galaxies, as classified in the BPT diagram. For all other galaxies, either AGN or non emission line galaxies, the SFRs are inferred from D4000-SFR relation (e.g. Kauffmann et al. 2003a). The SFR estimates based on H$\alpha$ flux, are corrected for dust extinction on the basis of the Balmer decrement. All SFR measures are corrected for the fiber aperture following the approach proposed by Salim et al. (2007). The stellar masses and SFRs are computed by assuming a Kroupa IMF and are adjusted to a Chabrier IMF for consistency with the other datasets. The accuracy of the H$\alpha$ and D4000 SFR indicators is performed and discussed in the Appendix.

\subsubsection{The WISE dataset}
The GALEX-SDSS-WISE Legacy Catalog (GSWLC, Salim et al. 2016, hereafter S16) is obtained by cross-matching the SDSS spectroscopic catalog with the GALEX UV and WISE database in addition to SDSS and 2MASS photometric information. It contains physical properties of ~700,000 galaxies with SDSS redshifts at $0.01<z<0.30$. GSWLC contains galaxies within GALEX footprint, regardless of a UV detection, altogether covering 90\% of SDSS. We use, in particular, the subsample with medium-deep GALEX observations of $\sim$ 1500 s exposure (GSWLC-M), which covers 49\% of the SDSS area. This is done to combine relatively deep UV observations and high statistics. Indeed, the "deep" catalog (GSWLC-D) samples only 7\% of the SDSS area, resulting in a rather poor statistics. GSWLC  utilizes WISE observations  at  22$\mu$m  (WISE channel  W4)  to  determine  SFRs  independently of the UV/optical SED fitting. The depth of WISE observations over the  sky  is  not  uniform,  but  is  still  much more uniform than GALEX depth, and essentially covers the entire  sky  without  gaps. The average $5\sigma$ depth in the W4 channel if 5.4 mJy.

The mid-IR SFRs in GSWLC are estimated from  the  total  IR  luminosity  (8-1000 $\mu$m) by interpolating the luminosity-dependent IR templates of Chary \& Elbaz (2001) so that they match the 22 $\mu$m flux.  The IR luminosities ($L_{IR}$) are tested using Dale \& Helou (2002) templates, where the IR SED shape-luminosity  dependence  is  imposed  from  empirically calibrated relations of Marcillac et al. (2006). The agreement is excellent with a scatter of 0.02 dex. To obtain mid-IR SFRs from IR luminosity, S16 use a simple conversion given by Kennicutt (1998), adjusted to Chabrier IMF using the 1.58 conversion factor (S07):
\begin{equation}
log SFR = log(L_{IR})-9.996
\end{equation}
where SFR is in $M_{\odot}/yr$ and the $L_{IR}$ in $L_{\odot}$. 
The GSWLC provides also an estimate of stellar masses, SFRs and dust attenuations derived via SED fitting from the UV to the mid-IR data. The SED fitting is performed using the state-of-the-art UV/optical SED fitting technique code CIGALE (Noll et al. 2009). A comparison between SFRs derived from SED fitting and mid- and far-infrared derived SFRs is provided in Appendix. Despite a general agreement, the SFRs based on the SED fitting tend to be underestimated with respect to the WISE SFR, in particular in the region well above the local MS locus. This results in a very large scatter of 0.4 dex with respect to the 1 to 1 relation. 

Similarly to the GSWLC, also Chang et al. (2015) provide stellar masses and SFRs for a sample of 1 Million galaxies drawn from the SDSS New York University Value-Added Galaxy Catalog (NYU-VAGC, Blanton et al. 2005;Adelman-McCarthy et al. 2008;Padmanabhan et al. 2008)
and crossed-match with WISE data. However, as shown in the Appendix, the SFRs based on the SED fitting derived with MAGPHYS (da Cunha et al. 2008, 2012) from the UV to the W3 or W4 WISE channel at 12 and 22 $\mu$m, respectively, tend to be underestimated with respect to the H$\alpha$-based and the SPIRE-based SFR. Thus, we use, in the further analysis, the WISE and SED fitting SFR of GSWLC.

\subsubsection{The H-ATLAS {\it{Herschel}}/SPIRE dataset}
Bourne et al. (2016) provide the Data Release I catalog of multiwavelength associations to the H-ATLAS sources detected in the 5 {\it{Herschel}} bands at 100 and 160 $\mu$m with PACS and 250, 350 and 500 $\mu$m with SPIRE over an area of $\sim$ 150 deg$^2$. The IR sample is described in Valiante et al. (2016). Namely, the 250 $\mu$m detections are used as priors for the source extraction in the remaining {\it{Herschel}} bands. Thus, the H-ATLAS catalog is a 250 $\mu$m selected catalog. The Valiante et al. (2016) catalog includes 120230 sources in total, with 113995, 46209 and 11011 sources detected at $>4\sigma$ at 250, 350 and 500 $\mu$m, where the 1$\sigma$ level is 7.4, 9.4 and 10.2, mJy, respectively. Bourne et al. (2016) provide a cross-match with SDSS, GALEX, 2MASS and WISE photometry and with the SDSS and GAMA spectroscopic catalogs. We use the subsample of H-ATLAS DRI catalog with spectroscopic counterpart at $z < 0.085$. 

The total IR luminosity is estimated from the far-IR data as follows. We use all the available far infrared data-points, with the addition of the WISE 22 $\mu$m data point, when available, to compute the IR luminosities integrating the best spectral energy distribution (SED) template in the range 8-1000\,$\mu$m. To this aim we use two different sets of templates to check the model dependence. We use the Main Sequence (MS) and starburst (SB) templates of Elbaz et al. (2011) and the set of Magdis et al. (2014). All templates provide infrared luminosities extremely consistent to each other with a rms of 0.05 dex (see Appendix).
We refer to this {\it{Herschel}} far-infrared luminosity based SFRs as the most accurate measure of the galaxy SF activity. This is used in the Appendix to check the reliability of most of the other SFR indicators.

\subsection{SFR indicators and samples}

Each local galaxy sample described above presents different biases due to the combination of different selection effects and SFR indicators. To investigate such biases we check the reliability of the different SFR indicators and methods by comparing different SFR measures with the far-infrared derived SFR, based on all the available mid and far infrared data-points. All the plots of such comparison are shown in Appendix. Here we report the main results of such analysis. As shown in the Appendix, the {\it{Herschel}} based SFR are in very good agreement with the H$\alpha$ based SFR of the MPA-JHU catalog, with a 0.2 dex scatter (left panel of Fig. \ref{app3} in Appendix). The same agreement is observed between the mid-infrared WISE SFR of S16 and the {\it{Herschel}} based SFR, with a scatter of 0.18 dex around the 1 to 1 relation (Fig. \ref{app6} in Appendix). 

A much larger discrepancy is observed with the D4000 based SFR of the MPA-JHU sample. In the majority of the cases the D4000 based SFR estimate is heavily underestimated with respect to the far-infrared based SFR (left panel of Fig. \ref{app4} in Appendix). The unreliability of the D4000 SFR estimates is confirmed also when comparing the SFRs based on the mid-infrared WISE SFR at 22 $\mu$m, which sample a larger range of SFR with respect to the far-infrared H-ATLAS dataset (right panel of Fig. \ref{app4} in Appendix). Also in this case the D4000 based SFRs appear to be on average 0.7 dex below the 1 to 1 relation with a quite large scatter of 0.4 dex. On the contrary, when compared with the SED fitting SFR of S16 below the MS, the D4000 based SFR tend to provide a 1 to 2 dex overestimated SFR (Fig. \ref{app4} Appendix), showing to be highly unreliable also in the quiescence region. 

\begin{figure}
	\includegraphics[width=\columnwidth]{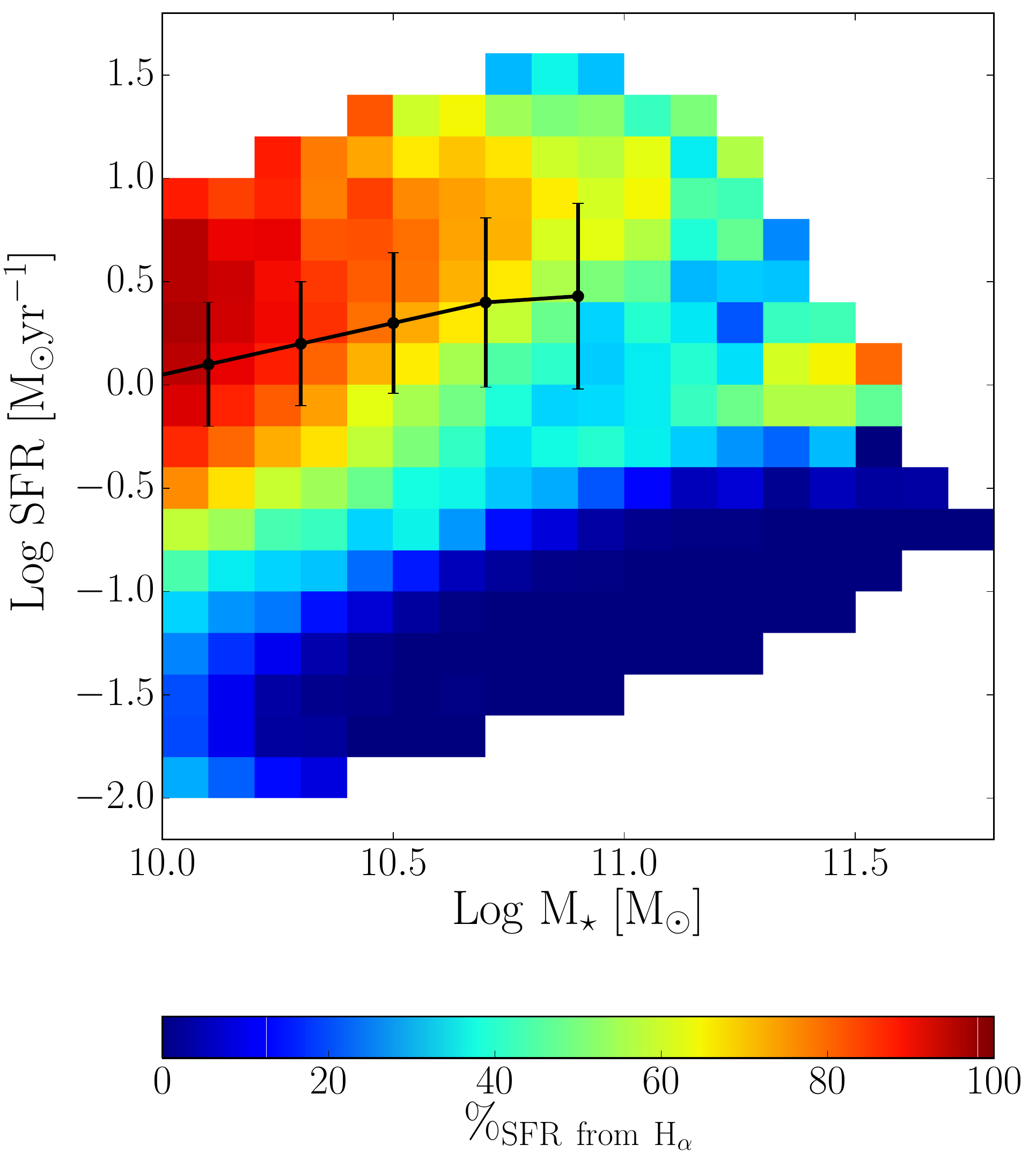}
\caption{Distribution of MPA-JHU galaxies in the log(SFR)-log(M*) plane color-coded as a function of percentage of galaxies with SFR derived from dust corrected H$\alpha$ emission, in bin of SFR and stellar mass. The peak of the MS distribution at different stellar mass bins in the log(SFR)-log(M*) plane is over-plotted (connected points) to show the peak of the SFR distribution in the Ms region. The error bars shows the dispersion of the relation around the peak.}
\label{sfr_sdss}
\end{figure}

As shown in Fig. \ref{sfr_sdss}, the H$\alpha$ based SFRs of the MPA-JHU catalog sample very well the region down to to 2$\sigma$ below the MS of RP15 up to stellar masses of $10^{10.6}$ $M_{\odot}$. Above this threshold, the unreliable D4000 SFR estimates dominate the lower envelope of the MS and above $10^{10.8}$ $M_{\odot}$ they account for most of the SFR of the MS. It is, then, clear that the MPA-JHU SFR estimates can not be used to  study the location of the MS and its shape above $\sim 10^{10.6}$ $M_{\odot}$.

To overcome this problem we use the calibration proposed by Oemler et al. (2017). As shown in the Appendix, such correction, based mainly on the galaxy inclination and the rest-frame $NUV-g$ color, where $NUV$ is the GALEX near-uv filter and $g$ is the SDSS g band, is able to correct most of the effects mentioned above, as clearly outlined also in the appendix of Oemler at al. (2017).  In the further analysis, we will consider only the SFRs corrected according to the Oemler at al. (2017) calibration (see the Appendix for more details).

The SFRs based on the SED fitting of S16 correlate with the {\it{Herschel}} based SFRs with a larger scatter of 0.28 dex with respect to the WISE SFRs of the same catalog (Fig. \ref{app6} in Appendix). We point out, however, that as shown in Appendix in Fig. \ref{app7}, the SFR based on SED fitting of the UV and optical range tend to be underestimated up to 0.5 dex at higher level of SFR. Indeed, we see a clear anti-correlation between the ratio of the SED fitting SFR and the far-infrared SFR as a function of the distance from the MS. For this exercise we use the Main Sequence of RP15 up to $10^{10.6}$ $M_{\odot}$. This indicates, as expected, that the SFR based on UV$+$optical data is not capable of capturing the SF activity of the most star forming and dusty objects well above the MS. The ratio of the two SFR estimates is, instead, around 1 for galaxies on and below the MS.

While the SFR based on WISE data sample mainly the MS region but only partially the lower envelope, the combination of the SED fitting based SFR below the MS and the WISE based SFR above and on the MS, provides the most reliable set of SFR to study the location and the shape of the MS in the local Universe.  Instead, due to the rather shallow flux limit at 250 $\mu$m, the H-ATLAS galaxy sample is located mainly in the upper envelope of the MS and provide only a partial sampling of the relation.

In order to investigate the possible biases induced by the different sets of SFR estimates and selections, we perform the analysis of the SFR distribution in the MS locus in the same way, separately, with different galaxy samples. We discuss, then, consistency, discrepancies and biases. In particular we use:
a) the MPA-JHU catalog corrected with the calibration of Oemeler et al (2017, hereafter 017), defined by its SFR indicators as "H$\alpha+$D4000 O17" sample, b) the S16 sample with 22 $\mu$m WISE SFR, when available, and SFR from SED fitting for all WISE undetected galaxies (hereafter "WISE$+$SED fit" sample), c) the subsample of S16 limited to WISE detected sources at 22 $\mu$m, complemented with galaxies classified as star forming in the BPT diagram and with SFR derived from dust corrected H$\alpha$ in the MPA-JHU catalog (hereafter "WISE$+$H$\alpha$" sample), d) the S16 sample with SFR based purely on SED fitting (hereafter "SED fit"), and e) and the H-ATLAS sample (hereafter "SPIRE" sample). The sample a) is used to study the distribution of the SFR around the MS below $10^{10.8}$ $M_{\odot}$ in the upper envelope and at the peak of the MS, which is mostly sampled by the H$\alpha$ based SFR estimates (Fig. \ref{sfr_sdss}). The sample a) and  b) are able to cover also the lower envelope and the quiescence region of the log(SFR)-log(M*) plane and are used to investigate the SFR distribution also well below the MS. The sample c) is based on the most reliable SFR indicators available in the local Universe and it is used to study accurately the shape of the SFR distribution in the MS region. The sample d) is used to check the biases due to the UV selection and the SED fitting technique. The e) sample is limited to the shallow H-ATLAS survey, and it is used only to check the shape of upper envelope of the SFR distribution retrieved with the other samples. In all cases we apply a redshift cut at $z=0.085$ in order to ensure mass and SFR completeness down to $10^{10} M_{\odot}$ (see also Peng et al. 2010). 

All SFR and stellar masses are converted to a Chabrier IMF, when necessary, for consistency.

\begin{figure*}
\includegraphics[width=18cm,height=18cm]{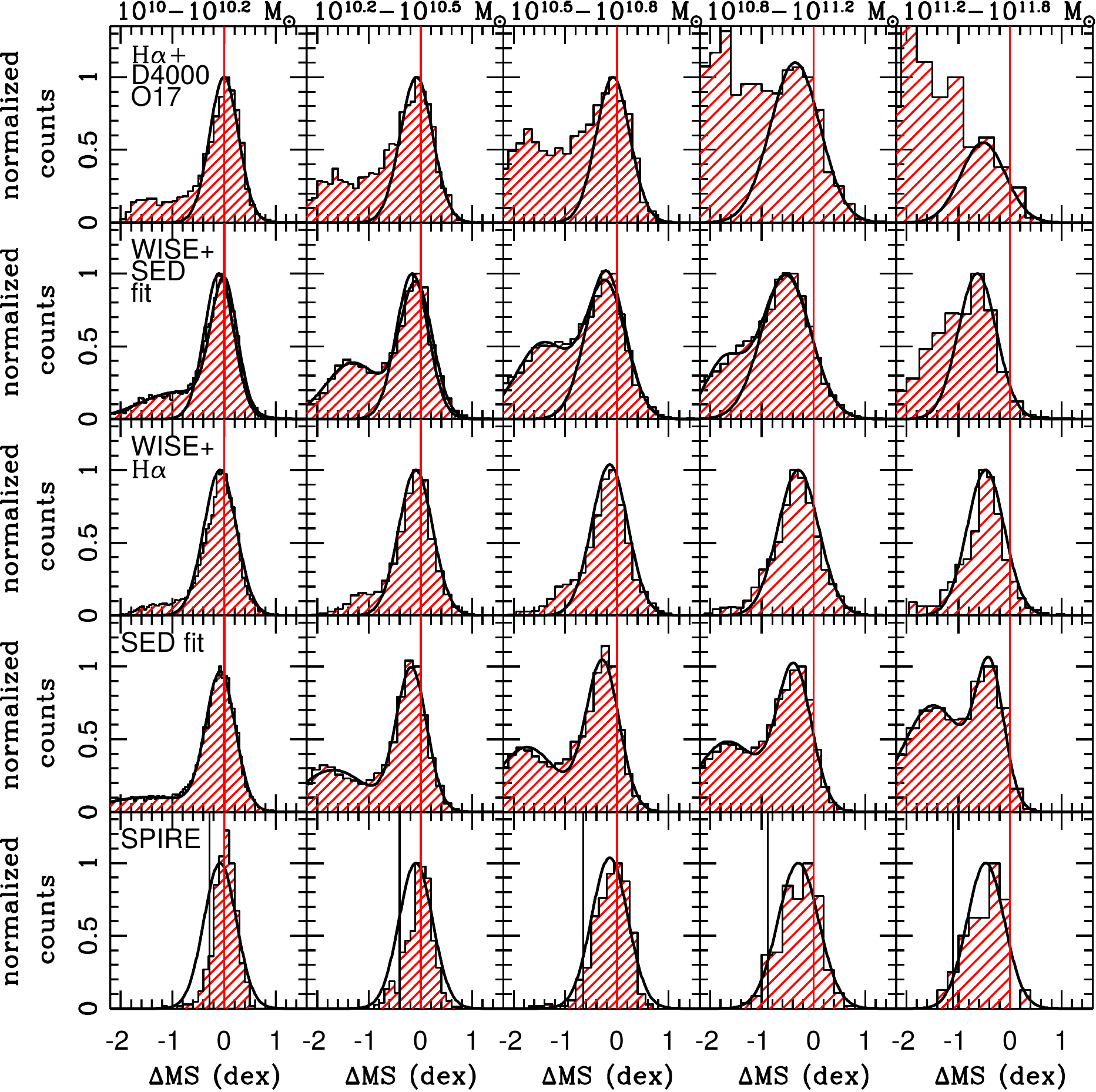}
\caption{Distribution of the residual $\Delta{MS}$ around the MS in several stellar mass bins in the local Universe (red shaded histogram). The vertical red line, in all panels, shows the $\Delta{MS}=0$ point corresponding to the MS location given by the relation of RP15. Each row shows the distribution based on different samples. From the top to the bottom we show the distribution of the residuals based on a) the MPA-JHU SFR (corrected for the O17 calibration) and stellar mass estimates ("H$\alpha+D4000$ O17"), b) the S16 sample with SFR based on WISE 22 $\mu$m data-point, when available, and SFR based on SED fitting elsewhere ("WISE$+$SED fit") c) the S16 subsample with SFR based on WISE 22 $\mu$m data-point, when available, and SFR based on dust corrected H$\alpha$ flux from MPA-JHU catalog for BPT classified SF galaxies ("WISE$+$H$\alpha$"), d) the S16 sample with SFR based on SED fitting only ("SED fit"), e) the H-ALTLAS 250 $\mu$m selected sample with SFR derived from far-infrared data ("SPIRE). In the bottom row the black vertical line show the luminosity limit corresponding to the 5$\sigma$ limit of the 250$\mu$ selected sample of Valiante et al. (2016). The solid curve in each panel shows the best fit normal distribution in the log-log space. For the "WISE$+$SED fit", and "SED fit" samples we show also the best fit curve of the whole distribution including the quiescence region. In all panels the distribution is normalized to the peak of the MS distribution.}
\label{fig3}
\end{figure*}

\subsection{Sample completeness}

In order to limit the analysis to a stellar mass complete sample, we cut each sample at $z < 0.085$ and stellar masses of $10^{10}$ $M_{\odot}$.

For the "H$\alpha+D4000$ O17" sample, the completeness in SFR is the same as for the mass, because a measure of the SFR is provided for each galaxy either from the H${\alpha}$ emission or from the D4000-SFR scaling relation of Salim et al. (2007), corrected for the O17 calibration see previous subsections for details). For the S16 "WISE$+$SED fit" and "SED fit" samples, the completeness in SFR for galaxies above our stellar mass threshold is the same, because a measure of the SFR is always provided at least via SED fitting. For the S16 "WISE$+$H$\alpha$" sample the completeness is dictated mainly by the completeness limit of the WISE catalog in the W4 channel, whose sources dominate in number the sample. To be conservative, we indicate such threshold, corresponding to the SFR at the 5$\sigma$ WISE flux limit, as our completeness limit. However, we point out that complementing the S16 WISE subsample with the SF galaxies of the MPA-JHU sample populates the lower envelope of the MS, because the H$\alpha$ based SFR mostly sample the MS well below the peak of the distribution at least up to $10^{10.6}$ $M_{\odot}$, as shown in Fig. \ref{sfr_sdss}. 

Above this mass threshold, however, the H$\alpha$ based SFR is available only for a small percentage of galaxies, in particular in the lower envelope of the MS. Thus, the SFR distribution of the S16 "WISE$+$H$\alpha$" sample in the high mass range might be biased against faint sources in the far-infrared (less dusty objects). The "SPIRE" sample, which is purely far-infrared selected, shares the same problem. Its completeness limit is set to the SFR corresponding to the 5$\sigma$ limit at 250 $\mu$m. 

\section{The Definition of the local MS}

In this section we measure the location of the MS in the local Universe at $z < 0.085$, where the SDSS spectroscopic sample offers an exquisite statistics for such purpose. 

\subsection{The SFR distribution in the log(SFR)-log(M*) plane}

For each sample we analyze the distribution of the galaxies SFR in a fine grid of stellar mass bins (0.2 dex wide in the $10^{10}-10^{11} M_{\odot}$ stellar mass range). In order to properly find the location of the MS and define the SFR distribution around it, we fit the SFR distribution with a log-normal component. We stress that we exclude from the fitting procedure the lower envelope of the MS, which could bias either the location or the dispersion of the best fit log-normal. The fit is limited to the region down to 0.3 dex below the MS, so to avoid any bias due to a possible excess of galaxies in the valley between the MS and quiescence region. 

In the following analysis, we study the shape of the distribution of the residuals $\Delta{MS}=log(SFR_{gal})-log(SFR_{MS})$ given by the distance from the SFR of the individual galaxy ($SFR_{gal}$) and the SFR on the MS at the mass of the galaxy ($SFR_{MS}$). In particular, we use as reference the relation of RP15. This is obtained as the ridge line connecting the peak of the SFR distribution in the log(SFR)-log(M*) plane in the stellar mass range $10^8-10^{10.5}$ $M_{\odot}$, where the H$\alpha$ SFRs of the MPA-JHU sample dominate in number the MS region (see previous section). Thus, we consider it an accurate estimate of the MS location in the low stellar mass regime. The best fit is a power law with $log(SFR) \propto log(M*)^{\alpha}$, with $\alpha=0.76$ and scatter $\sigma \sim 0.3$. This approach allows us to verify if the MS is bending at higher stellar masses with respect to the best fit relation of the lower stellar mass regime.

\begin{figure}
\includegraphics[width=\columnwidth]{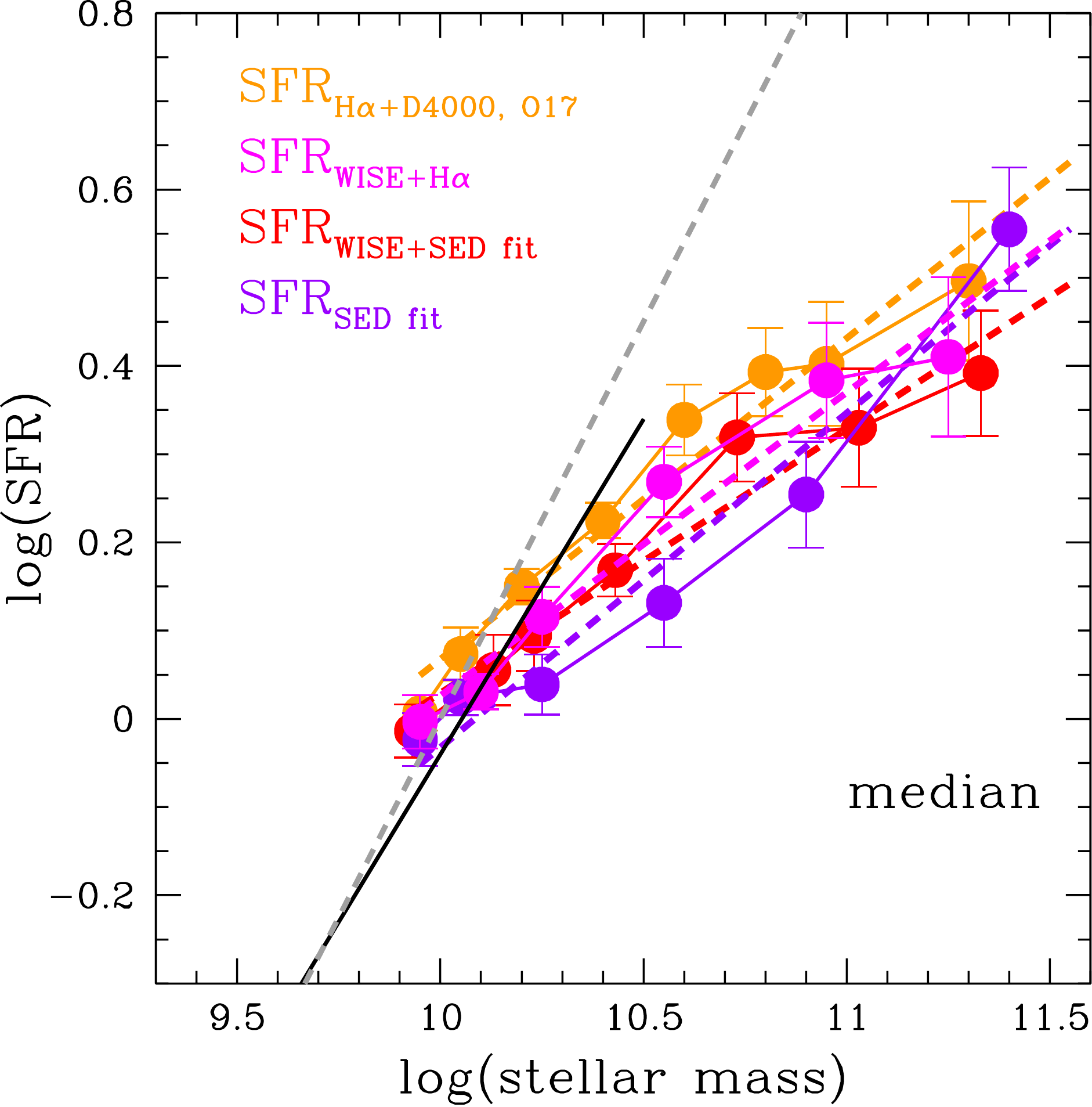}
\caption{Location of the MS based on the {\it{median}} of the best fit log-normal distribution for the different samples. The dashed lines of different color show the best fit linear regression on each sample. The color code of points and lines is indicated in the figure. The black line indicates the relation of RP15 up to $10^{10.5}$ $M_{\odot}$, and the gray line shows the relation of Peng et al. (2010).}
\label{ms_indi}
\end{figure}

Fig. \ref{fig3} shows the distribution of the residuals $\Delta{MS}$ for the different galaxy samples considered here. In all cases the upper envelope is well fitted by a log-normal distribution (the black Gaussian in logarithmic scale in each panel). 

The lower envelope of the MS, instead, deviates, as expected, from a log-normal distribution with increasing significance towards higher masses, in particular in the "H$\alpha+$D4000 O17", the "WISE$+$SED fit" and "SED fit" samples. However, the shape of the distribution below the MS strongly depends on the SFR indicator used. In the "H$\alpha+$D4000 O17" sample the lower envelope and the high mass end of the MS are sampled mainly by the D4000 indicator, corrected according to the O17 calibration. In "WISE$+$SED fit" and "SED fit" samples, the valley and the peak in the quiescence region are sampled by SFRs derived through SED fitting, mainly driven by the GALEX rest-frame UV flux.

The red vertical line in each panel shows the $\Delta{MS}=0$ location corresponding to the value of the MS of RP15. It is apparent that the peak of the $\Delta{MS}$ distribution is not consistent with 0 in all samples towards high stellar masses. The significance of the deviation, however, depends on the SFR indicator used for the analysis. Obviously the "H$\alpha+$D4000 O17" distribution is the most consistent with the RP15 log-linear relation, because it is based on the a similar dataset. Nevertheless, for all other samples, the peak of the distribution falls below 0 at least above $10^{10.5}$ $M_{\odot}$, suggesting that the MS location is bending towards lower values of SFR with respect to the relation of the low stellar mass regime.

The last row of Fig. \ref{fig3} shows the residual distribution based on the "SPIRE" data, selected at 250 $\mu$m in the H-ATLAS sample. At lower masses the H-ATLAS survey does not sample the whole MS due to the shallow flux limit. We compare the H-ATLAS distribution with the WISE best fit after volume correcting the best fit normalization. A Kolmogorov-Smirnov (KS) test reveals that the WISE best fit log-normal distribution is a good fit also for the far-infrared selected sample, in particular at high masses, where the MS is fully sampled.

\subsection{The MS location}

In order to define the location of the MS, we use the derived best fit log-normal distributions in each sample. We use three different indicators: the mean, the median and the mode of the log-normal distribution retrieved by fitting the upper envelope of the MS. If $\mu$ and $\sigma$ are the mean and dispersion of the normal distribution of the logarithmic variable $log(SFR)$, the mean, median and mode of the log-normal distribution of the linear variable, SFR, are: 
\begin{equation}
mean=10^{\mu+{\sigma}^2/2}
\end{equation}
\begin{equation}
median=10^{\mu}
\end{equation}
\begin{equation}
mode=10^{\mu-\sigma^2}
\end{equation}
In particular, the median SFR provides the {\it{geometric}} mean of the linear distribution. The median SFR coincides with the mode of the Gaussian distribution in log(SFR), used for instance in Ilbert et al. (2015). The {\it{arithmetic}} mean SFR is instead used, by construction, in all the stacking analysis of the MS, in particular at high redshift (e.g. Whitaker et al. 2014, Rodighiero et al. 2010, 2014, Schreiber et al. 2015). The {\it{geometric}} and {\it{arithmetic}} mean of a log-normal distribution always differ, with the former being smaller than the latter. Thus, some caution must be used in comparing results obtained with different statistical indicators. The mode of the log-normal SFR distribution has never been used in the literature and we will not consider it in the further analysis.

\begin{table*}
\caption{Best fit Linear regression in the log(SFR)-log(M*) of the MS relation. The parameter {\it{a}} and {\it{b}} indicate the slope and zero point of the best fit linear regression for each sample and for the median and mean MS, respectively. \label{TAB:fit}}
\begin{center}
\begin{tabular}{lcccc}
    \hline
    \hline \\[-2.5mm]

 &    median     &   &    mean	& \\	       
    \hline
           &    a   &  b  &                a   &  b   \\
              \hline
 H$\alpha+D4000 O17$ &  $0.36\pm 0.03$ & $-3.57\pm 0.31$  &  $0.41\pm 0.03$ & $-4.09\pm 0.34$ \\ 
 WISE$+$ SED fit  &  $0.30\pm 0.03$ & $-2.97\pm 0.32$  &  $0.35\pm 0.03$ & $-3.43\pm 0.35$ \\
 SED fit         &  $0.38\pm 0.04$ & $-3.83\pm 0.44$  &  $0.38\pm 0.04$ & $-3.78\pm 0.42$ \\
 WISE$+$H$\alpha$  &  $0.34\pm 0.03$ & $-3.41\pm 0.38$  &  $0.36\pm 0.04$ & $-3.57\pm 0.43$ \\

   \hline
\end{tabular}
\end{center}
\end{table*}

Fig. \ref{ms_indi} shows the MS location according to median indicator for each sample. All samples agree within 1.5$\sigma$. They all show a clear peak in the MS locus at any stellar mass and with a significant bending with respect to the RP15 relation obtained at lower stellar masses. Table 1 reports the best fit parameters of the linear regression performed in the log-log space on the median and mean SFR points, estimated from the best log-normal fit, shown in Fig. \ref{ms_indi} for each sample. In all cases the slope of the relation is in the range 0.30-0.38$\pm$0.02, which is much flatter than the 0.76 slope reported at lower stellar masses by RP15. 

\begin{figure}
\includegraphics[width=\columnwidth]{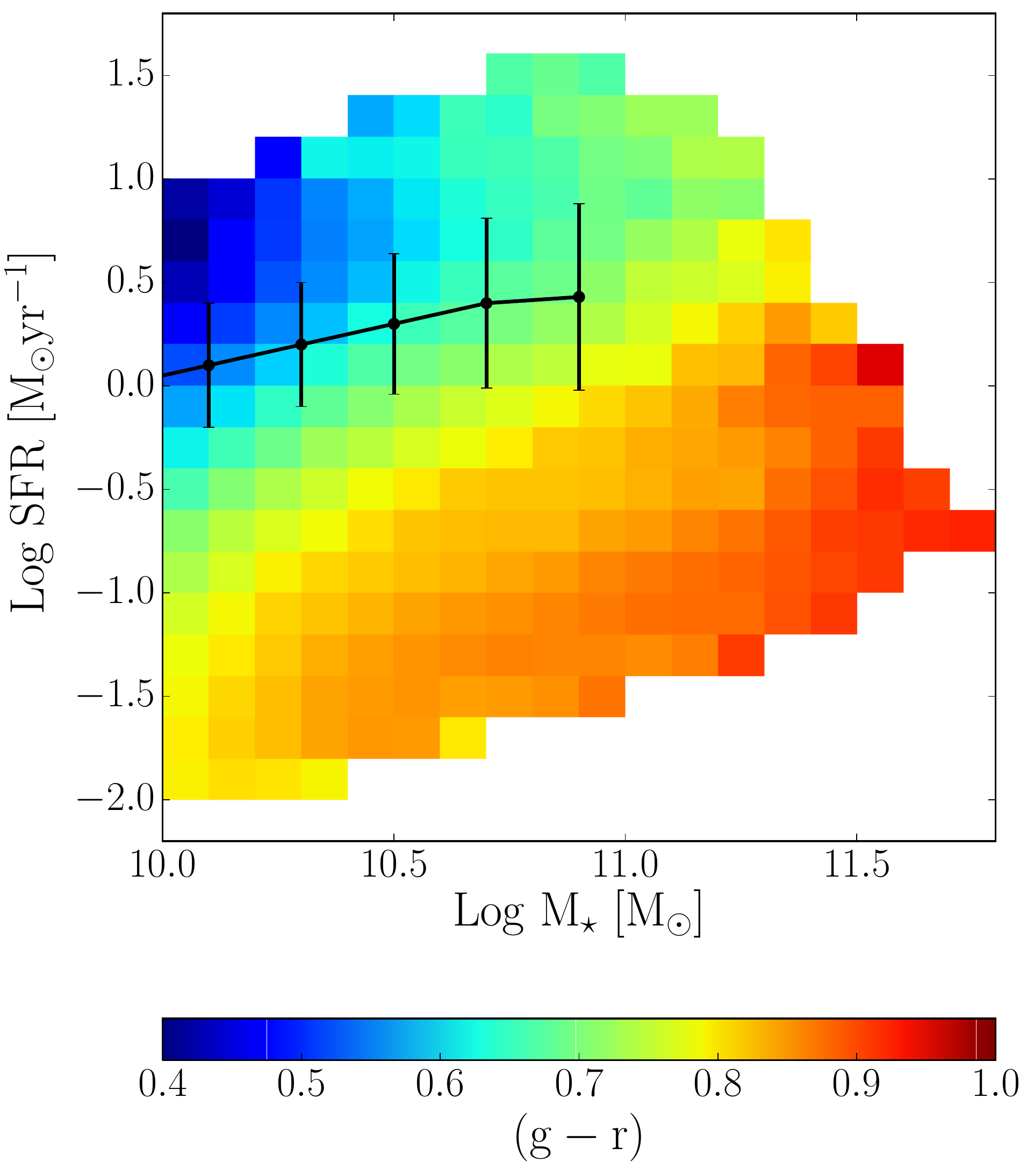}
\caption{Distribution of galaxies in the log(SFR)-log(M*) plane color-coded as a function of the mean g-r color in bin of SFR and stellar mass. The mode of the MS at different stellar mass bins is over-plotted (connected points). The error bars shows the dispersion of the relation around the mode. }
\label{color_cut}
\end{figure}

The median SFR of the the S16 "SED fit" sample distribution lies below the other relations almost at any mass bin, but it is still consistent within 1.5$\sigma$. We ascribe this discrepancy to the systematic underestimation of the SFR from SED fitting in the upper envelope of the MS, as shown in the Appendix (Fig. \ref{app5}). Indeed, as visible in the histograms of Fig. \ref{fig3}, such region is much less populated in this sample than in the WISE$+$H$\alpha$ sample, with the results of an overall underestimation of the log(SFR) distribution peak. We plot also the relation of Peng et al. (2010) based on liner regression between SFR and stellar mass for blue galaxies in the log-log space. We point out that such selection changes the SFR distribution around the MS, because, as shown in Fig. \ref{color_cut}, galaxies are getting redder not only across but also along the MS. The selection of only blue galaxies excludes most of the systems populating the high mass end of the MS and favors galaxies in the upper envelope of the relation, leading to a steeper relation.

\begin{figure}
\includegraphics[width=\columnwidth]{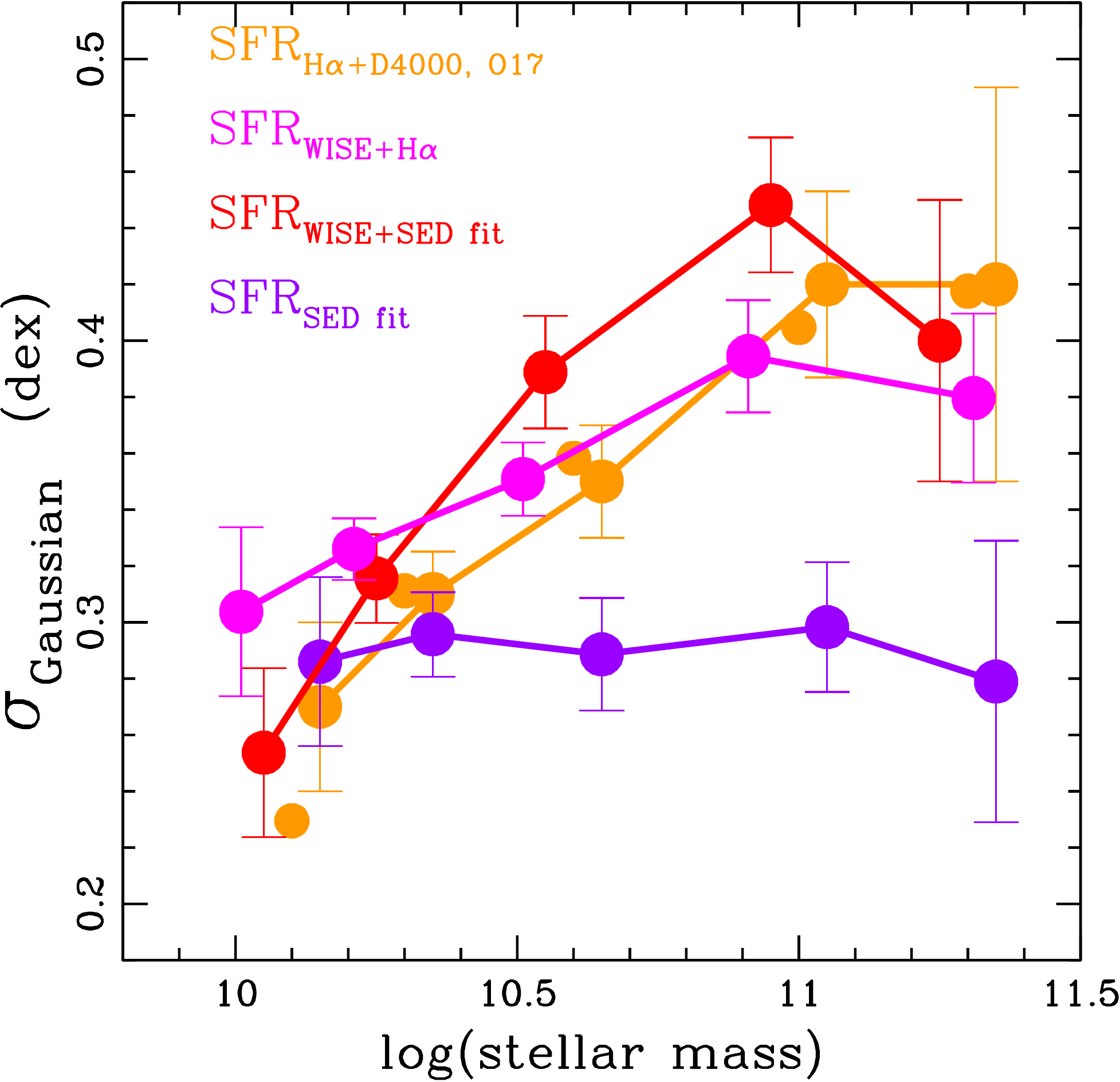}
\caption{Dispersion $\sigma$ of the best fit normal distribution of the log(SFR) variable in the log-log plane for different samples. The color code is indicated in the figure. }
\label{scatter}
\end{figure}

The location of the mean, instead, depends on the value of the dispersion of the normal distribution of the logarithmic variable. As shown in Fig. \ref{scatter}, the dispersion of the normal distribution in the logarithmic space, is increasing as a function of stellar mass in all samples but in the S16 "SED fit" sample. Also this effect is related to the overall underestimation of the SFR from SED fitting in the upper envelope, which squeezes the distribution to a fix scatter of $\sim$ 0.3 dex. Instead, all other samples indicate a scatter ranging from $\sim$ 0.3 dex at $10^{10}$ $M_{\odot}$ to $\sim$ 0.4 dex at $10^{11}$ $M_{\odot}$. This is found also by using the "WISE$+$H$\alpha$ sample and it is confirmed by the far-infrared based SFR of the H-ATLAS sample. We stress that the increase of scatter is purely driven  by the galaxies in the upper envelope of the MS because the log-normal fit is performed only down to 0.3 dex below the peak. Hence there is no bias induced by the excess of galaxies in the valley between MS and quiescent region. The increase of the scatter implies that the MS relation provided by the mean SFR is steeper than the relation based on the median of the distribution, as indicated in Table 1. As for the MS based on the median SFR, the slope of the relation remains flatter than the 0.76 slope reported at lower stellar masses, lying in the range 0.35-0.41$\pm$0.02.

As previously mentioned, the number of galaxies in excess with respect to the log-normal component in the lower envelope of the MS depends strongly on the SFR indicator. Indeed, it appears to be quite different also between the "WISE$+$SED fit" and the "SED fit" samples, that differ for the inclusion of the WISE based SFR. Such excess is likely responsible for the different level of flattening reported by different works in the literature in the local Universe MS. Indeed, if a running mean or median is used to define the location of the MS for such non log-normal distribution, the excess of galaxies in the valley moves the value of arithmetic mean and median well below the mean and the median of the log-normal distribution. The exact location of such values is strongly dependent on how the SF galaxy sample is defined in first place and on the SFR indicator used. 

\begin{figure}
\includegraphics[width=\columnwidth]{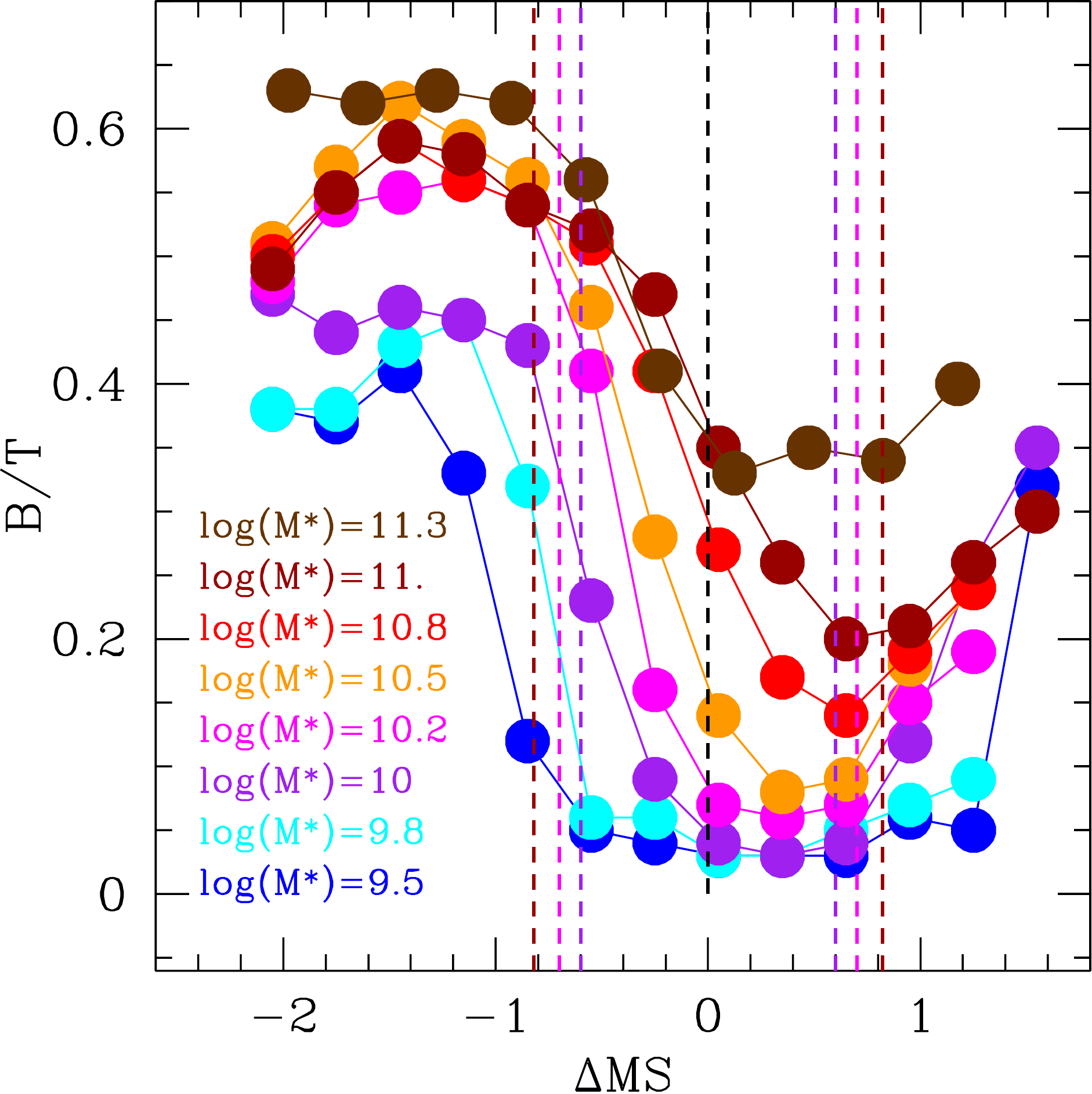}
\caption{Mean B/T as retrieved from Simard et al. (2011) as a function of the distance from the MS in several stellar mass bins. The points and lines are color coded as a function of the stellar mass bin as indicated in the figure. The dashed black lines at $\Delta{MS}=0$ indicate the location of the MS based on the median SFR. The other dashed line indicate the 2$\sigma$ interval around the MS at the stellar mass bin of the same color.}
\label{morpho}
\end{figure}

\begin{figure}
\includegraphics[width=\columnwidth]{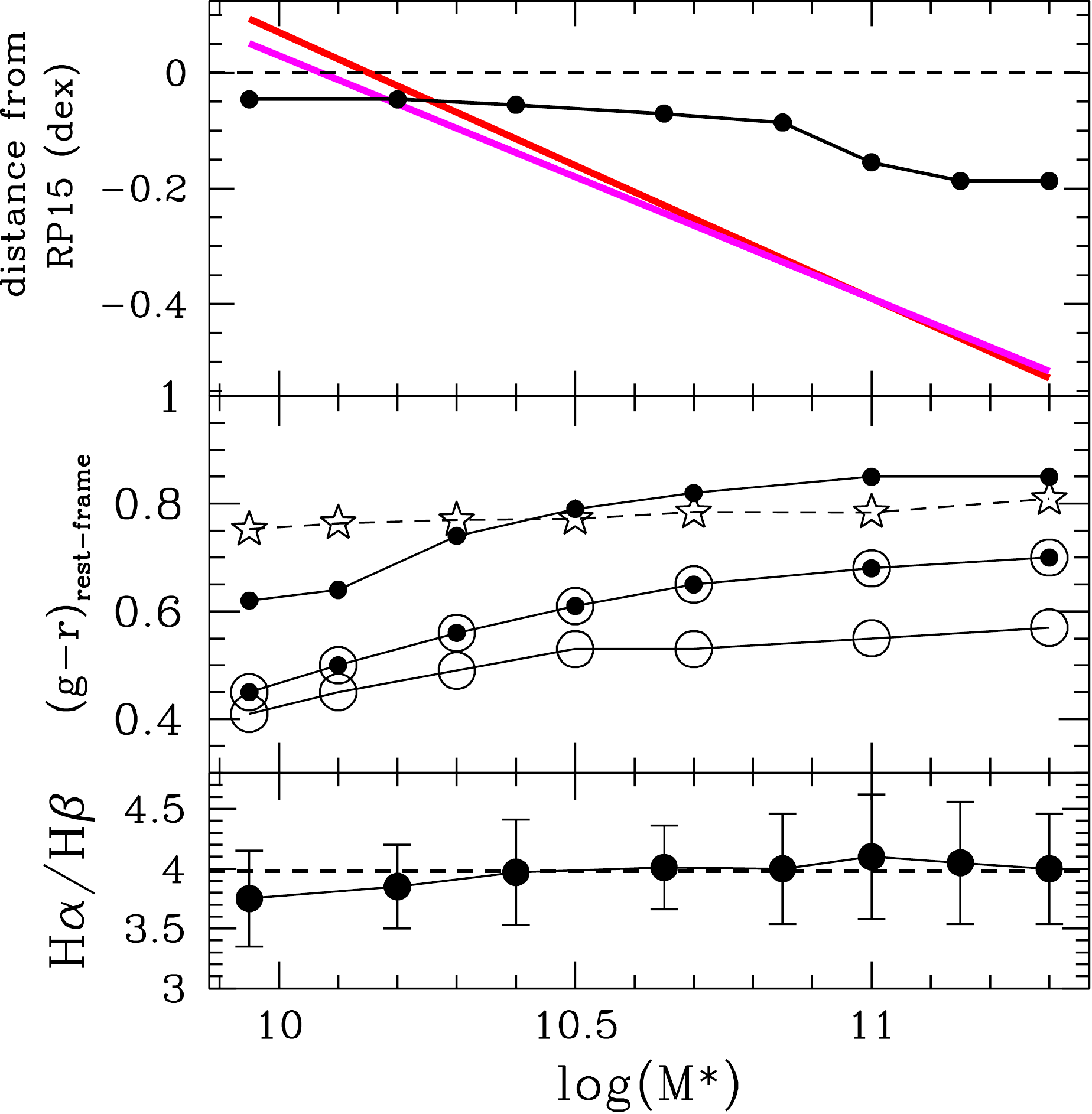}
\caption{{\it{Upper panel}}: Distance in dex from the power law relation of RP15 derived for galaxies at stellar masses below $10^{10.5}$ $M_{\odot}$ of the best fit power law retrieved in this work for the "WISE$+$H$\alpha$" (magenta line) and the "WISE$+$SED fit" (red line) samples, respectively. The black curve shows the RP15 MS relation reduced by a percentage equal to the mean B/T of the galaxies within 1$\sigma$ from the MS, under the assumption that the bending is due to the increase of the red and dead bulge component along the MS, as seen in Fig. \ref{morpho}. {\it{Middle panel}}: Mean (g-r) rest-frame color of Simard et al. (2011) matched to the S16 "WISE$+$SED fit" sample for galaxies within 1$\sigma$ from the median SFR of the MS in bins of stellar mass. The color is shown for the whole galaxy (empty points with central filled points), for the bulge component (black points) and for the disk component (empty points). The stars indicate the mean color of spheroidal galaxies in the same mass bin and located in the quiescence region at more than 1.5 dex below the MS. {\it{Bottom panel}}: Mean Balmer decrement (H$\alpha$/H$\beta$) of galaxies within 1$\sigma$ from the MS of the WISE$+$SED fit" sample. The H$\alpha$ and H$\beta$ fluxes are taken from the emission line flux catalog of Brinchmann et al. (2004). The error bars indicate the dispersion around the mean value.}
\label{color_dec}
\end{figure}

\section{Discussion}

We discuss here the two main findings of the paper which are {\it{i)}} the bending of the MS at high masses and {\it{ii)}} the increase of its scatter at high stellar masses. 

\subsubsection{The MS bending}
One of the most used explanation for the bending of the MS is the change of the galaxy morphology along it. Several works in the literature propose that the slope of the MS and its scatter are due to an increasing prominence of an inactive bulge coexisting with an actively star forming disk at increasing stellar masses (Wuyts et al. 2011, Whitaker et al. 2015, Abramson et al. 2014, Erfanianfar et al. 2016, Morselli et al. 2017). Abramson et al. (2014) show that, under the assumption that the bulge component is always inactive and that the SFR is purely due to the activity of the disk, one can remove part of the dependence of the sSFR on the stellar mass by normalizing SFR to the disk stellar mass. This flattens the slope of the sSFR-stellar mass relation to $-0.2$ with respect to what inferred from the RP15 relation. However, as shown in Morselli et al. (2017), such assumption is too simplistic. By measuring the mean B/T ratio of galaxies in the MS region derived with the MPA-JHU catalog, they show that the MS is sitting on the lowest value of the B/T, and that the mean B/T along the MS is increasing from 0.1 at $10^{10}$ $M_{\odot}$ to 0.4 at $10^{11}$ $M_{\odot}$. By matching the B/T estimates of Simard et al. (2011), as used in Morselli et al. (2017), with the S16 "WISE$+$SED fit" sample, we partially confirm the result. For this exercise we use the MS relation of RP15 below $10^{10}$ $M_{\odot}$ and our definition of the MS above the same mass limit, in order to take into account the bending of the relation at high masses. The mean B/T of galaxies sitting on the MS (at $\Delta{MS}=0$ in Fig. \ref{morpho}) increases from 0.1 at $\sim10^{9.5}$ $M_{\odot}$ to 0.35 above $\sim10^{11}$ $M_{\odot}$. The only difference is that the minimum B/T is not obtained on the MS but it is reached above the MS, with distance increasing as a function of the stellar mass (see fig. \ref{morpho}). This indicates that pure disk galaxies follow a relation steeper than the MS provided by the bulk of the star forming galaxy population, which at high masses is dominated by intermediate type morphology systems, as already seen in Salmi et al. (2016). This is also in agreement with the recent findings of Belfiore et al. (2018) based on ManGA data. Indeed, despite the limited statistics of the current MaNGA sample with respect to the whole SDSS spectroscopic or the S16 samples, Belfiore et al. (2018) find that the MS is consistent with the PR15 steep relation only when it is limited to galaxies dominated by star formation at all radii, as pure disk galaxies tend to be according to their colors. However, the MaNGA high mass end of the MS tend to be equally populated by completely star forming galaxies and systems with a LIER-like central region (cLIER systems) but star forming at larger galactocentric distances. If also cLIER systems are considered, the MaNGA MS does bend at high masses as found here. In addition, according to Belfiore et al. (2018) the pure SF galaxies tend to populate the upper envelope of the MS, while cLIER systems are located in the lower envelope, as clear from their Fig. n. 2. This would be consistent with the finding of Fig. \ref{morpho}, where the pure disk galaxies tend to have a much steeper MS than higher B/T systems. We will explore more in details this aspect in a dedicated paper.

However, the increase of the bulge component along the MS (from 10\% at $10^{10}$ to 35\% at $10^{11}$ $M_{\odot}$) does not suffice to explain the observed bending, as confirmed also by Abramson et al. (2014). As shown in the upper panel of Fig. \ref{color_dec}, if we reduce the extrapolation above $10^{10.5}$ $M_{\odot}$ of the RP15 MS relation (holding at $< 10^{10.5}$ $M_{\odot}$) by a percentage equal to the mean B/T of galaxies on the MS (black curve in the figure), the reduced SFR is still much higher with respect to the observed SFR at the MS high mass end retrieved with the "WISE$+$H$\alpha$" (magenta line) and the "WISE$+$SED fit" (red line) samples. Instead, one should take into account, as shown again in Morselli et al. (2017), that also the galaxy disk is getting redder along the MS, further lowering the SFR level at high masses. The results is confirmed also when using the S16 "WISE$+$SED fit" sample rather than the MPA-JHU catalog as in Morselli et al. (2017). In the middle panel of Fig. \ref{color_dec} we show the rest frame (g-r) color of galaxies within 1$\sigma$ from the median SFR in the MS region. The color of the galaxies as a whole is getting redder along the MS. The disk is slightly bluer than the whole galaxy and it follows the same curve getting redder at high masses. The bulge component is redder at any mass with respect to the whole galaxy and the disk component. In addition, it gets redder as a function of mass more rapidly, reaching a plateau above $10^{10.5}$ $M_{\odot}$. Above this mass threshold the mean color of the bulge component in MS galaxies is consistent with the color of spheroidal quiescent galaxies at the same mass and at 1.5 dex below the MS (star symbols in the middle panel of Fig. \ref{color_dec}). 
The galaxy color is much closer to the disk color than to the bulge component, confirming that the disk is dominating over the bulge component at any mass along the MS. However the color excess between galaxy and disk increases from 0.05 dex at $10^{10}$ $M_{\odot}$ to 0.13 dex at $10^{11}$ $M_{\odot}$, showing the effect of a larger red and dead bulge component. 
In order to check if the redder color of the galaxy and disk component at higher masses is due to a larger dust content rather than a mean older stellar age, we plot in the bottom panel of Fig. \ref{color_dec} also the mean value of the Balmer decrement, H$\alpha$/H$\beta$, for the same sample of galaxies. Indeed, the Balmer decrement is a direct estimate of the amount of dust in galaxies and it is used to correct the H$\alpha$ flux for dust extinction (Brinchmann et al. 2004). The values of the H$\alpha$ and H$\beta$ fluxes are taken from the emission line flux catalog of Brinchmann et al. (2004) for all galaxies with H$\alpha$ and H$\beta$ flux SNR higher than 3. This allows to sample 80 to 90\% of the whole galaxy population at 1$\sigma$ from the MS at any mass. As shown in the panel, the Balmer decrement is rather constant along the MS for galaxies at stellar masses above $10^{10}$ $M_{\odot}$. Thus, we conclude that the reddening of the galaxy and disk color is not due to an increase of galaxy dust content but rather to an increase of the mean stellar population age.
This suggests that the flattening of the MS at high masses is due to the two effects: the complete quenching of the bulge component above $10^{10.5}$ $M_{\odot}$ and a reduced star formation activity of the disk component at increasing stellar mass. This is confirmed by the spatially resolved sSFR maps of the MaNGA sample of Belfiore et al. (2018). Indeed, they find that the average radial sSFR profile of MS galaxies exhibits a significant central depression towards higher masses, in particular above $10^{10.5}$ $M_{\odot}$. This is also accompanied by a general depression of the sSFR level at all radii towards higher masses. Indeed, Belfiore et al. (2018) provide also a measure of the sSFR within different galactocentric annulii (their Table 2 and Fig. 7). Despite the large scatter, in particular towards the high mass end where the MaNGA statistics is limited, the sSFR-stellar mass relation appears in all cases as a clear anti-correlation with average slope of $\alpha=-0.24$. In addition, the relation based on the sSFR integrated in the central region appears to be steeper than the one based on the sSFR measured around and beyond the effective radius. This confirms that the SF activity is progressively suppressed at all radii towards the high mass end, and, in particular, such suppression takes place more rapidly in the central than in the external region of the galaxy.

\begin{figure*}
\includegraphics[width=18cm,height=9cm]{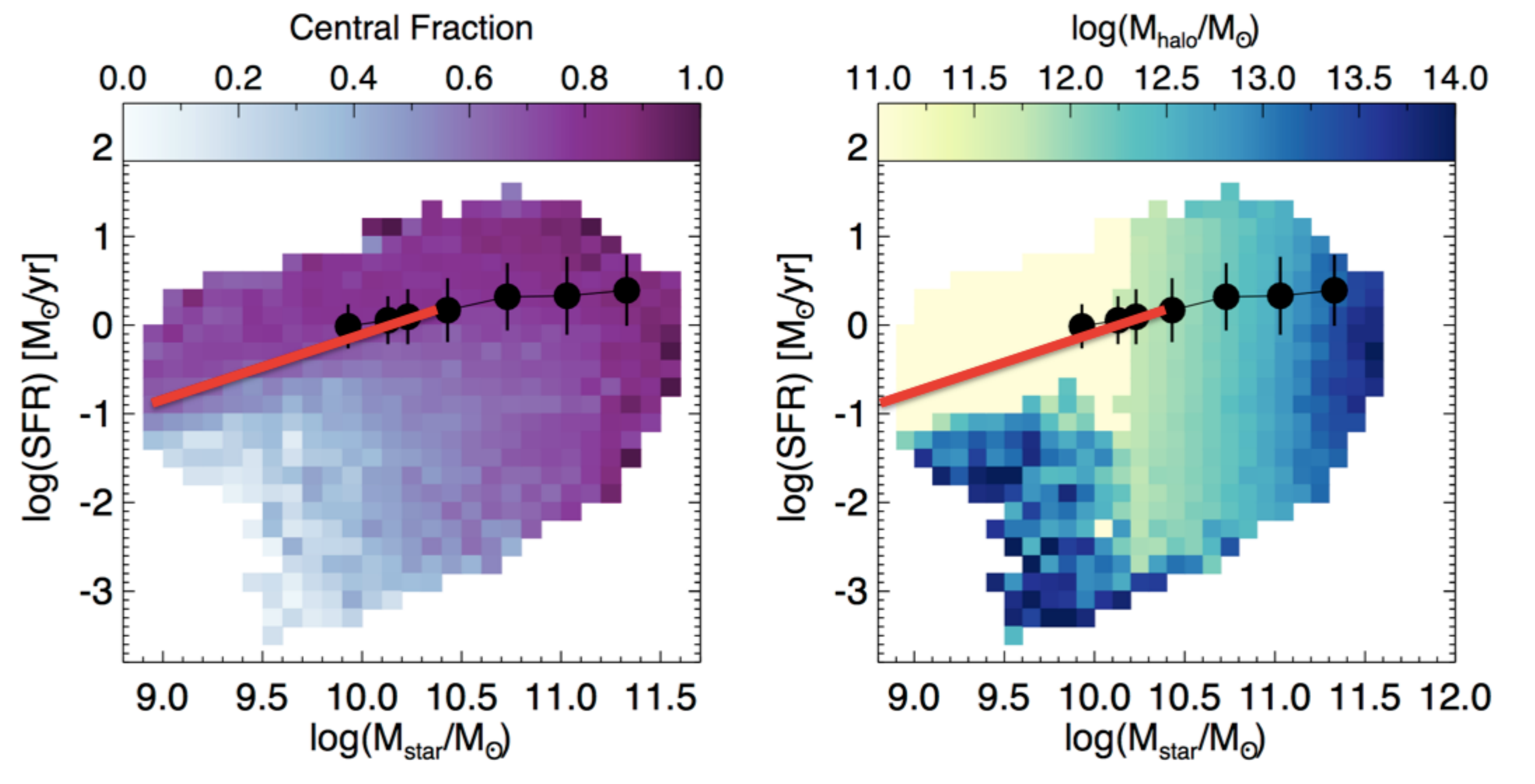}
\caption{{\it{Left panel}}: log(SFR)-log(M*) plane based on the S16 $WIDE+SED fit$ sample, color coded as a function of the faction of central galaxies identified in the host halo catalog of Yang et al. (2007). {\it{Right panel}}: same as the left panel but color coded as a function of the mean host halo mass in bin of SFR and stellar mass 0.1 dex $\times$ 0.1 dex wide. Also the host halo mass is retrieved by matching the S16 catalog with the halo mass catalog of Yang et al. (2007). The low mass regime (below $10^{10}$ $M_{\odot}$) is dominated by isolated galaxies, for which only an upper halo mass limit is provided in the Yang et al. (2007) catalog. This is why in this region it is not possible to observe any halo mass gradient as at higher masses. In both panels the black points indicate the location of the local MS based on the median indicator while the red solid line shows the relation of RP15 at lower masses.}
\label{env}
\end{figure*}

The quenching of the spheroidal component in nearly all semi-analytical models and hydrodynamical simulations is ascribed to the feedback of the central super-massive black hole (SMBH). There is a vast literature about the effects of the different implementations, either through powerful outflows (e.g.  Debuhr et al. 2011, 2012; Choi et al. 2012, 2014, 2015; Wurster \& Thacker 2013), or radio jets (Gaspari et al. 2011, 2014; Dubois et al. 2011, 2012, 2016; Meece et al. 2017; Schaye et al. 2015; Weinberger et al. 2017). However, while from the theoretical point of view there is consensus, a firm observational evidence is still lacking, in particular when it comes to the disk component. Indeed, if the effect of the SMBH feedback is recognized to have a local effect in the central galaxy region, it is still quite debated whether such effect can be extended on the kpc scale of the disk component (see for instance Martin et al. 2012, Rubin et al. 2010; Cicone et al. 2016; Fiore et al. 2017; Harrison et al. 2014; Brusa et al. 2015; Cresci et al. 2015; Concas et al. 2017, 2018; Chisholm et al. 2015, 2016, 2017). 

Thus, what else can cause the decrease of SF activity of the disk in massive galaxies? In this respect the effect of the environment might play a role too. To test this possibility we match the S16 "WISE+SED fit" sample to the host halo mass catalog of Yang et al. (2007) based on SDSS data. As shown in the right panel of Fig. \ref{env}, the mean host halo mass in the log(SFR)-log(M*) plane is increasing along the MS with the stellar mass. This is due to the fact that the vast majority of galaxies are central galaxies (left panel of Fig. \ref{env}) and to the known correlation between the $M*_{central}/M_{halo}$ versus $M_{halo}$, where $M*_{central}$ is the stellar mass of the central galaxy and $M_{halo}$ is the total mass of the host halo (Behroozi et al. 2013, Yang et al. 2009). It is worth to notice that the region where the MS is mostly bending above $10^{10.5}$ $M_{\odot}$ corresponds to host halo mass of $\sim 10^{12-12.5}$ $M_{\odot}$. Such mass range is considered to be the mass threshold for the transition between a regime of cold to hot accretion (Keres et al. 2009; Dekel \& Binboin 2006). In low mass halos, below $\sim 10^{12-12.5}$ $M_{\odot}$, the central galaxy is fed with cold gas by the gas streams coming from the cosmic filaments. In high mass halos, instead, the gas filling up the halo volume is shock heated to the halo virial temperature during the gravitational collapse and the cold gas streams are no longer able to penetrate the halo and replenish the central galaxy (Keres et al 2009). Feldmann et al. (2017) in a fine scale numerical simulation of a volume large enough to sample also galaxy groups, outline that the evolution of group galaxies is driven by {\it{i)}} mergers, which lead to the bulge formation, {\it{ii)}} the suppression of gas inflow in the hot atmosphere and {\it{iii)}} ram pressure stripping. Since the MS is dominated by central galaxies, mergers and suppression of gas inflow must play the most important role. The first would explain the increase of the bulge mass towards high stellar masses in the galaxy group regime of the MS (above stellar masses of $10^{10.5}$ $M_{\odot}$ and halo masses of $\sim 10^{12-12.5}$ $M_{\odot}$). The second would lead to the a lack of cold gas supply and a progressive lowering of the SF activity in the external region, as observed in this work and in Belfiore et al. (2018). We argue that ram pressure stripping might play a fundamental role for the low mass satellite. Indeed, Fig. \ref{env} shows clearly that low mass satellite galaxies of massive halos tend to be segregated below the MS, towards the quiescence region. This is in agreement with the findings of Oemler et al. (2017), who explain with a toy model the distribution of passive and quiescent galaxies below the MS. Indeed, by using a simple model for disk evolution based on the observed dependence of star formation on gas content in local galaxies, and assuming simple histories of cold gas inflows, they show that the evolution of galaxies away from the MS can be attributed to the depletion of gas due to the star formation after a cutoff of gas inflow.

However, we point out that the gas starvation induced by the gravitational heating in massive halos is not sufficient to guarantee the shut-down of the SF activity in the central galaxy. Indeed the cooling time of the hot gas in massive groups and clusters of galaxies, although long, is shorter than the Hubble time. Thus, the hot gas might cool down and lead to the formation of cooling flows able to replenish the galaxy of cold gas and trigger the SF process. In this respect, all simulations advocate once again the effect of the central SMBH to prevent the cooling of the gas by dumping large quantity of energy into the circum-galactic medium through radio jets and lobes (see Croton et al. 2006; Shaye et al. 2015; Genel et al. 2016).

\subsubsection{The increase of the MS scatter as a function of mass}

In this section we analyse the possible causes of the increase of the scatter as a function of stellar mass. Such increase is observed also by Ilbert et al. (2015) on a completely different dataset. Indeed, they use {\it{Spitzer}} MIPS mir-infrared and {\it{Herschel}} PACS far-infrared data in the COSMOS field up to redshift $\sim$1.4 to perform a very similar analysis.

As for the bending of the MS, much of the scatter of the relation is usually ascribed to the large range of morphological types of galaxies across the MS. Whitaker et al. (2015) show that the mean Sersic index of MS galaxies at intermediate and high redshift in the Candels fields is increasing not only along the MS, as shown in the previous section, but also across it from the starburst region towards the lower MS envelope. Morselli et al. (2017) confirms that in the local Universe, there is a large spread of B/T across the MS. However, the mean B/T increases above and below the MS. By matching the S16 "WISE$+$ SED fit" sample with the Simard et al. (2011) catalog in Fig. \ref{morpho}, we show that this is confirmed also with more accurate SFR estimates than the D4000 based SFR used in Morselli et al. (2017). We will not discuss here the increase of the B/T towards the star-burst region as it is largely discussed in Morselli et al. (2017). We only point out here that such behavior is confirmed by spatially resolved H$\alpha$ maps of galaxies in MaNGA (Ellison et al. 2018, Belfiore et al. 2018 and Guo et al. 2018) and SAMI data (Medling et al. 2018). 

However, we point out here that such spread in morphological types across the MS is not increasing as a function of the stellar mass above $10^{10}$ $M_{\odot}$. The mean B/T is increasing as a function of stellar mass but the dispersion around it for galaxies within 2$\sigma$ from the median SFR in the MS region is $0.12-0.15$ without any dependence on the stellar mass (see fig. \ref{morpho}). Thus, even in this case the change in morphological type mix across the MS is not sufficient to explain the increase in scatter as a function of mass. 

Ilbert et al. (2015) ascribe, instead, such aspect to the larger spread in star formation histories of massive galaxies with respect to the low mass counterparts. In this respect, Oemler et al. (2017) point out that the peak of the Main Sequence should be determined by the age of the galaxies and by mean value $SFR(t)/<SFR>$. If the bulk of the galaxies are roughly coeval, then only the range of shape of the star formation histories $SFR(t)$ will set the scatter of the relation, at fixed stellar mass. The particular shape of $SFR(t)$ is not relavant.  Oemler et al. (2017) clearly show that any star formation histories that are not completely random will produce a correlation of SFR with mass with the observed scatter. Since the majority of the MS galaxies are statistically central galaxies, their SF and accretion history is bound to the merger history of their host halo. Hirschmann et al. (2013) nicely show that the merger three of a group-sized halo of mass $\sim 10^{13}$ $M_{\odot}$ is much more complex with respect to a $\sim 10^{11}$ $M_{\odot}$ halo. Indeed, in massive halos the frequency of merger and accretion episodes is much higher than in low mass halos. Such increased complexity might lead to a much larger range of possible evolutionary patterns for massive central galaxies, thus to a larger spread in $SFR(t)/<SFR>$ and so to a larger scatter of the MS towards the high mass end.

\section{Summary and conclusions}

We summarize here briefly the main findings of the paper. We study the Main Sequence of star forming galaxies in the local Universe by analyzing the SFR distribution in the log(SFR)-log(M*) plane. To this aim, we use different SFR indicators, including SED fitting, H$\alpha$ derived SFR and mid- and far-infrared derived SFR, in order to take into account selection effects and biases. 

In the local Universe, the SFR distribution in the MS region is well fitted by a log-normal distribution only in the upper envelope of the MS, consistently for all SFR indicators. At lower SFR, the distribution shows a clear excess of galaxies with respect to the log-normal distribution. Such excess increases as a function of stellar mass, but its significance depends strongly on the SFR indicator. Similarly, the dispersion of the log-normal distribution is not constant but it increases as a function of mass. The most robust SFR indicators, such as the dust corrected H$\alpha$ and the infrared derived SFR, agree in finding a dispersion ranging from $\sim 0.3$ dex at $10^{10} M_{\odot}$ to $\sim$ 0.4 dex at $10^{11} M_{\odot}$, as found with a very similar analysis by Ilbert et al. (2015) up to $z\sim 1.4$. Above $10^{11} M_{\odot}$, the location of the MS is very uncertain and it strongly depends on the SFR indicator used. 

We use different indicators to identify also the location of the MS, such as the median and the mean of the best log-normal SFR distribution in several stellar mass bins. For all SFR indicators, the MS relation flattens progressively at high stellar masses with respect to the relation found by RP15 at lower masses. The significance of such bending depends, though, on the SFR indicator and on the MS indicator used in the analysis. 

Because of this systematics, one must be cautious in comparing different results in the literature. Indeed, several slopes of the MS has been estimated ranging from 0.6 to 1.2, with or without bending at the high mass end. This suggest that such discrepancies might be due to the different indicators used either to define the MS location or the estimate the SFR, or galaxy sample selection effects. 

The analysis of the mean B/T in the MS region along and across the relation shows that the B/T increases from 0.1 to 0.35 from $10^{10}$ $M_{\odot}$ to $10^{11}$ $M_{\odot}$ is not sufficient to explain the bending of the MS at very high masses. In addition to the increase of the bulge component we observed also indication for a decrease of the SF activity of the disk along the MS. While the quiescence of the bulge component could be due to the effect of central SMBH feedback, we speculate that the lower SFR of the disk at high masses could be due, instead, to the gas starvation induced by the gravitational heating in massive halos. Indeed, we observe that above $10^{10.5}$ $M_{\odot}$, where the bending is most significant, the totality of the galaxies in the MS region are central galaxies of group and cluster sized halos. As suggested by Ilbert et al. (2015), the larger spread of merger threes and evolutionary paths of the group and cluster central galaxies could explain also the increase of the MS log-normal component as a function of the stellar mass. 

We conclude that the slope and the shape of the MS in the local Universe are dictated by the interplay between morphological transformation and environment and not simply by the stellar mass.

\section*{Acknowledgements}
This research was supported by the DFG cluster of excellence "Origin and Structure of the Universe" (www.universe-cluster.de). We thank L. Abramson for the extremely useful comments, that helped improving the manuscript. P.P. thanks also M. Sargent, E. Daddi and A. Renzini for the very useful discussion.








\appendix

\section{Comparison of different SFR indicators}

\begin{figure}
\includegraphics[width=0.9\columnwidth]{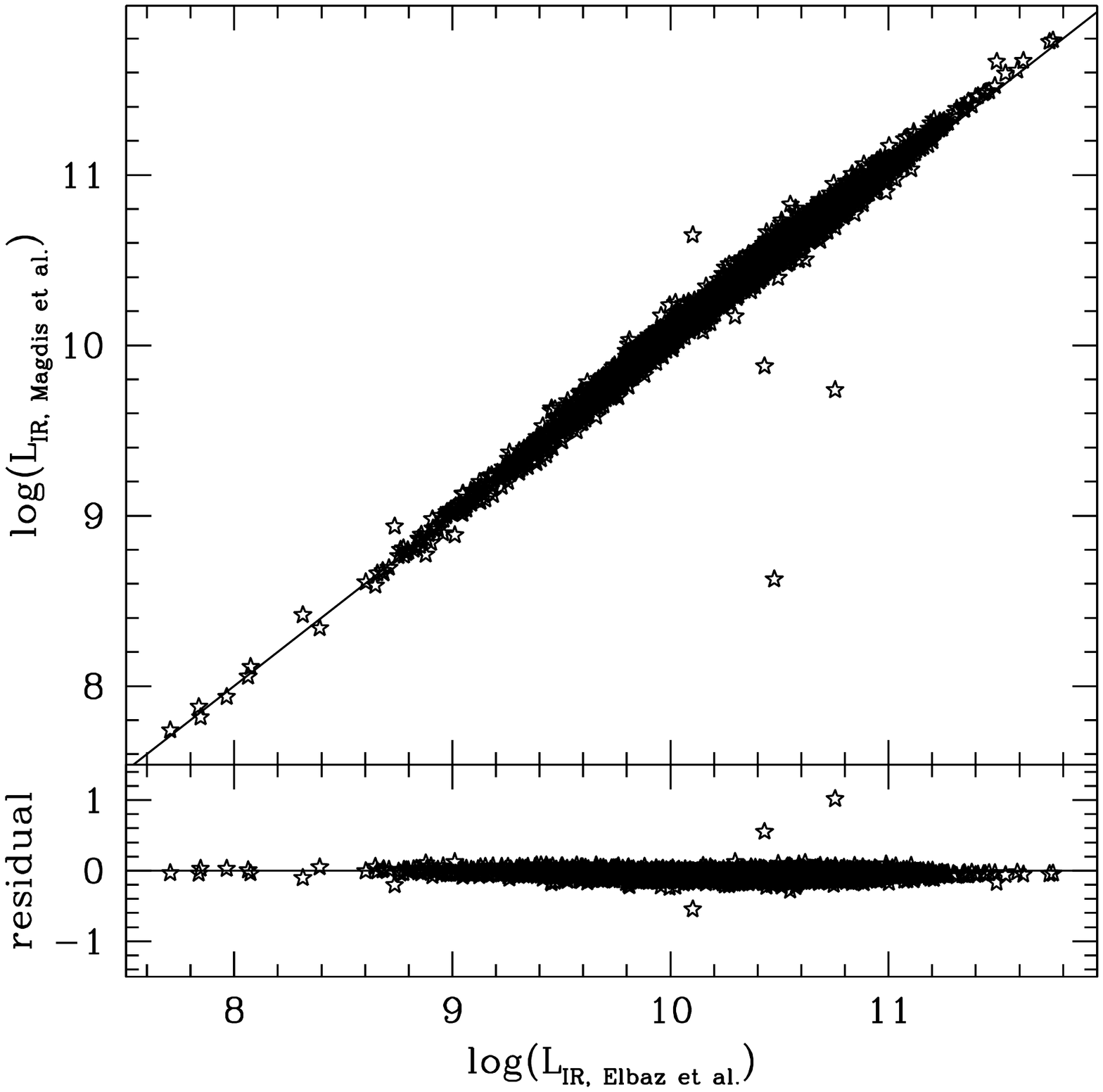}
\caption{Comparison of the far-infrared $L_{IR}$ derived in the low redshift sample with the Elbaz et al. (2011) MS and SB templates versus the $L_{IR}$ derived with the Magdis et al. (2014) templates. The $L_{IR}$ is obtained by integrating the template best fitting the WISE, PACS and SPIRE data from 8 to 1000 $\mu$m.}
\label{app1}
\end{figure}

\begin{figure}
\includegraphics[width=0.9\columnwidth]{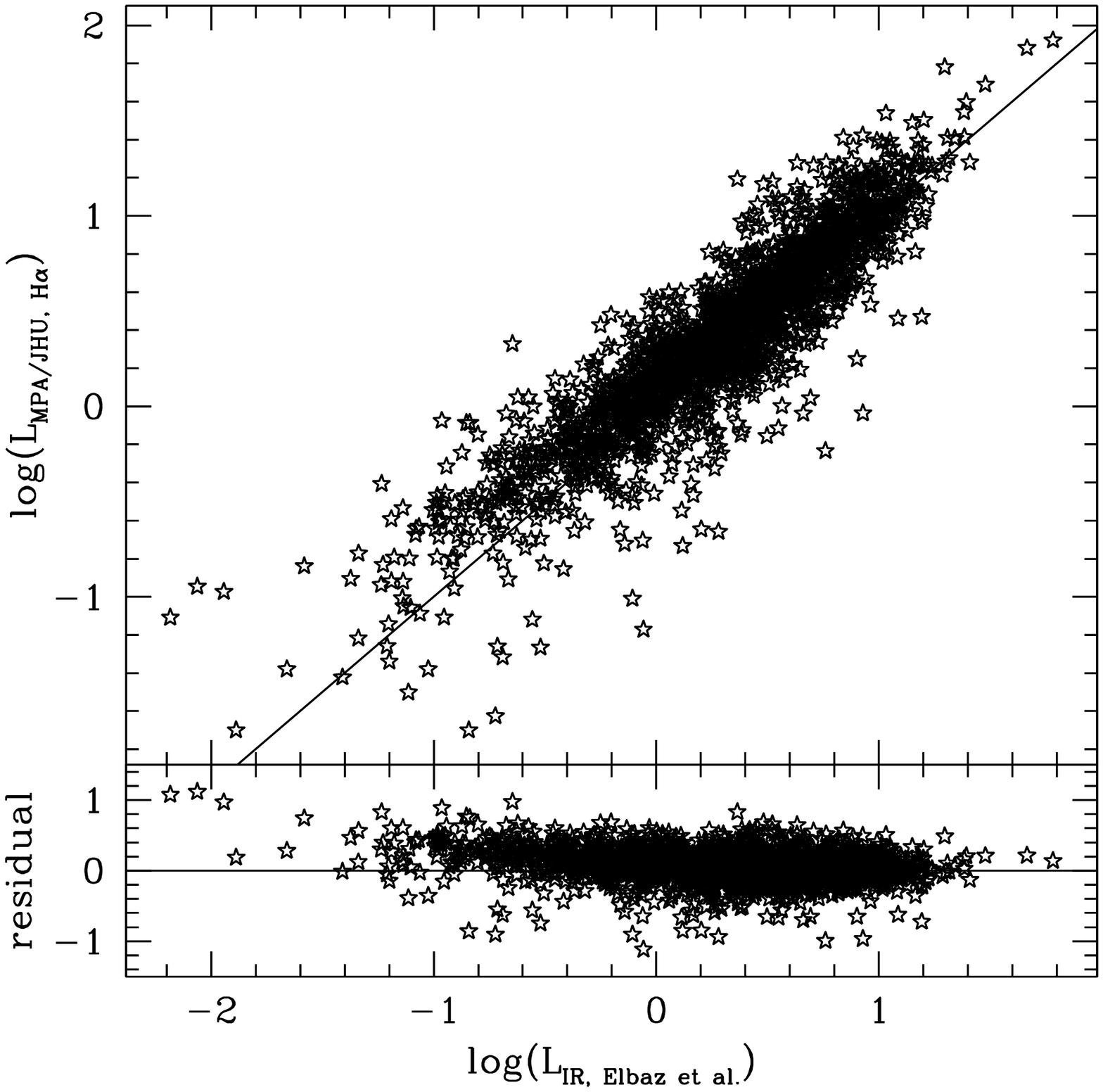}
\caption{Comparison of the SFR based on the far-infrared $L_{IR}$ derived in the low redshift sample with the Elbaz et al. (2011) MS and SB templates versus the dust corrected H$\alpha$ based SFR of the MPA-JHU catalog.}
\label{app3}
\end{figure}

In this section we compare different SFR indicators available for the local galaxy sample. In particular, we consider as the most robust SFR estimate the one based on the $L_{IR}$ measured by integrating from 8 to 1000 $\mu$m the best fit template of the H-ATLAS {{\it{Herschel}} SPIRE 250 $\mu$m data point and all available IR data points from 22 $\mu$m to 500 $\mu$m. Fig. \ref{app1} shows the comparison of such $L_{IR}$ obtained by using the Elbaz et al. (2011) and the Magdis et al. (2010) templates, respectively. The agreement is very good with an rms of 0.08 dex. 

\begin{figure*}
\includegraphics[width=0.9\columnwidth]{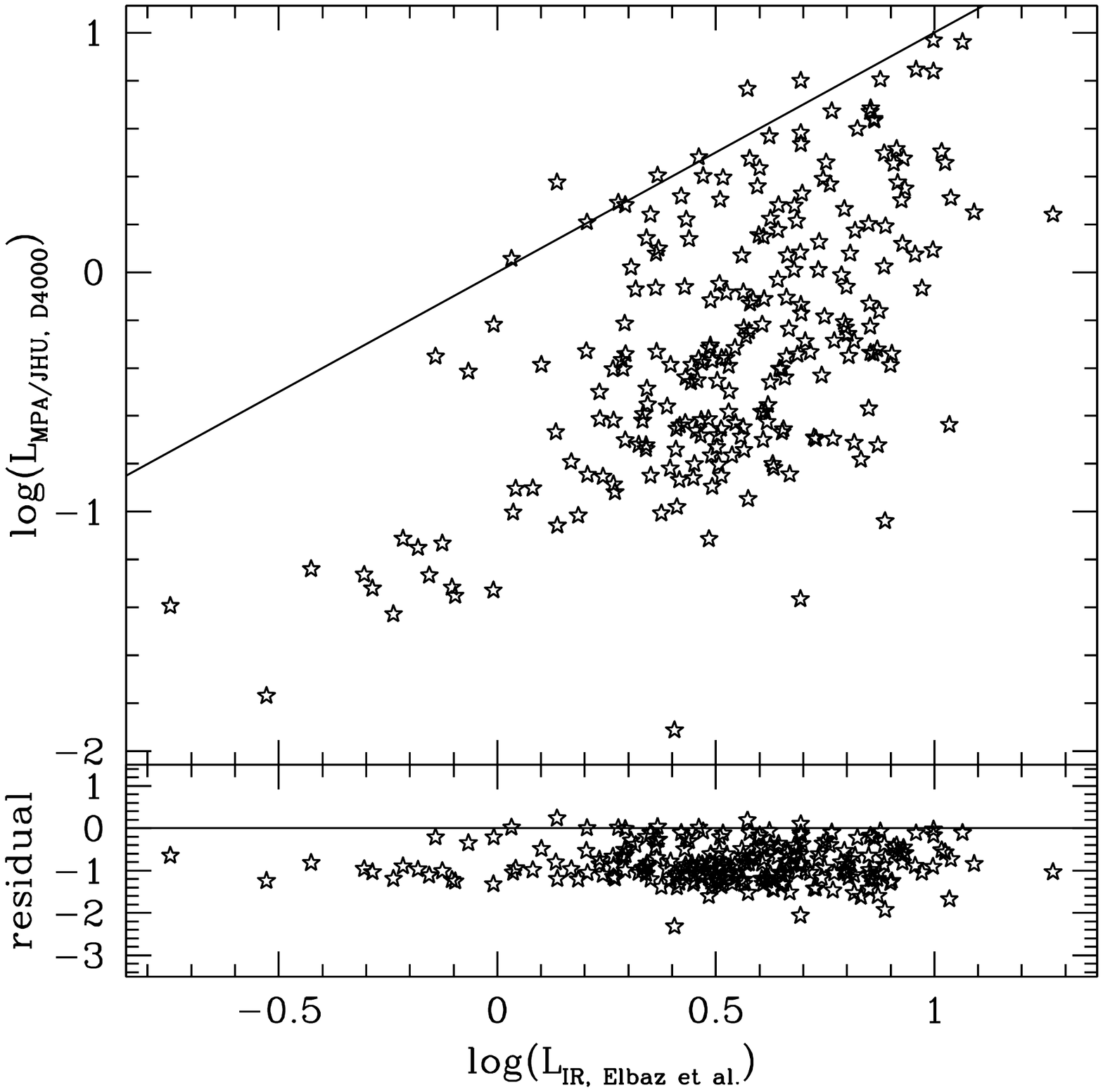}
\includegraphics[width=0.9\columnwidth]{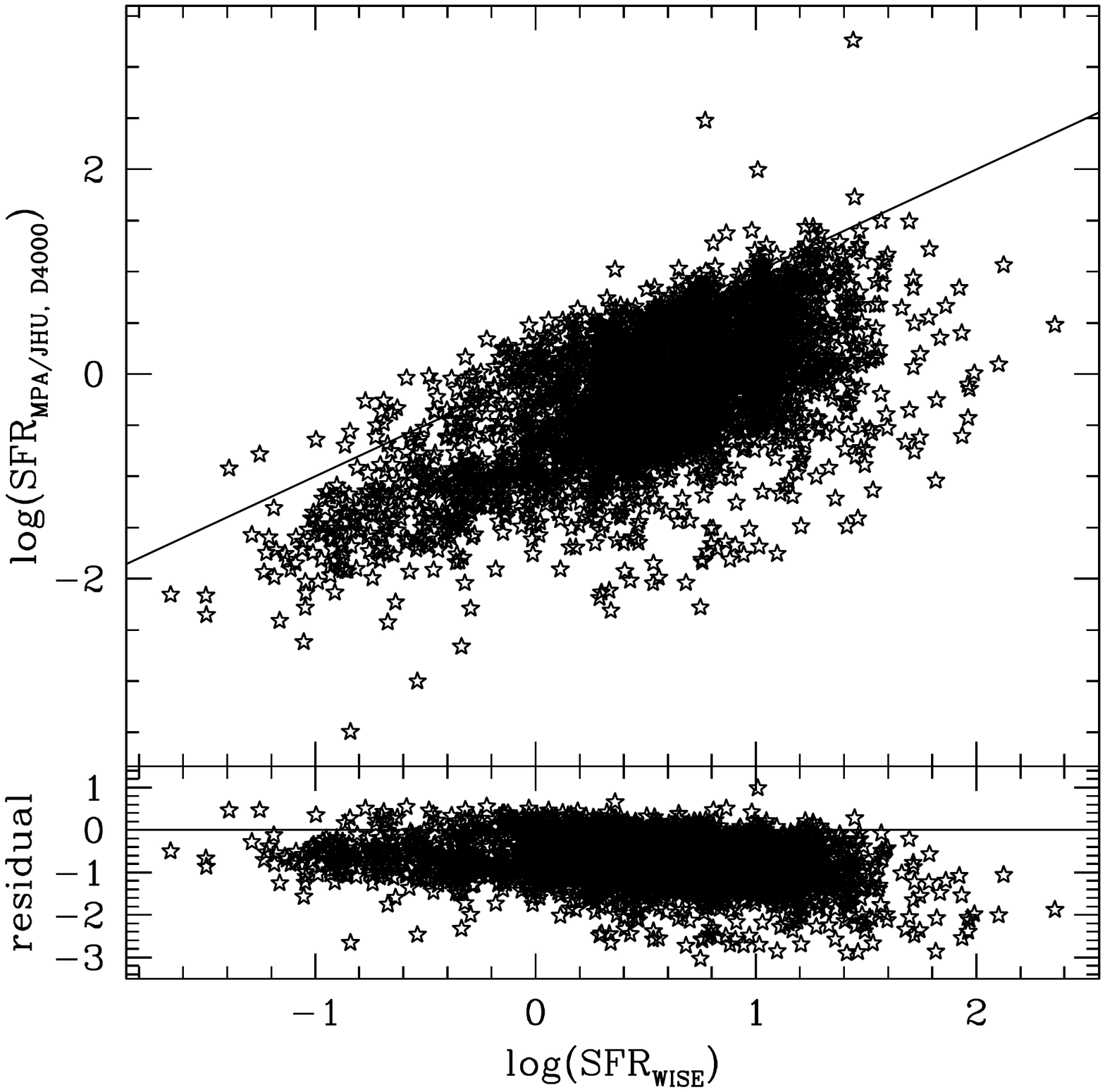}
\caption{{\it{Left panel}}:Comparison of the SFR based on the far-infrared $L_{IR}$ derived in the low redshift sample with the Elbaz et al. (2011) MS and SB templates versus the D4000 based SFR of the MPA-JHU catalog. {\it{Right panel}} Comparison of the SFR based on the mid-infrared WISE SFR of S16 versus the D4000 based SFR of the MPA-JHU catalog.}
\label{app4}
\end{figure*}

We keep as a reference the SFR based on the $L_{IR}$ derived with the Elbaz et al. (2011) templates in the further comparison. Fig. \ref{app3} show the comparison of the {{\it{Herschel}} based SFR with the H$\alpha$ based SFR of the MPA-JHU catalog (Brinchmann et al. 2004). The H$\alpha$ based SFR are in very good agreement with the IR SFR with a scatter of 0.23 dex. The D4000 based SFR are, instead, largely underestimated with respect to the IR based SFR (Fig. \ref{app4}). Such underestimation is confirmed also when comparing the D4000 derived SFRs with the WISE based SFRs of S16, which are sampling a larger SFR range with respect to the H-ATLAS sample. On the contrary, when compared with the SFRs based on SED fitting of S16, which sample mainly the quiescent region, the D4000-derived SFRs appear to be heavily overestimated at low SFRs (below 0.01 $M_{\odot}yr{-1}$) up to 2 orders of magnitude (Fig. \ref{app4a}). This is due to the fact that, for quiescent galaxies, the D4000 provides by construction only an upper limit to the actual level of star formation activity, imposed by the minimum value of the Balmer break. 
Oemler et al. (2017) propose a calibration of the H$\alpha$ and D4000-derived SFRs able to largely correct for such effects. The H$\alpha$ and D4000-derived SFRs are calibrated versus IR+UV based SFR with a correction based on the galaxy inclination and the $NUV-g$ rest-frame color, where $NUV$ is the GALEX NUV filter and $g$ is the SDSS g filter. We follow the calibration provided in Oemler et al. (2017). To this aim, use the Simard et al. (2011) catalog of bulge-dik decomposition to retrieve the galaxy inclination and the Bianchi et al. (2017) revised GALEX catalog to retrieve the NUV magnitude for the local SDSS sample. Fig. \ref{app4b} shows the corrected H$\alpha$ and D4000-derived SFRs  versus the WISE based SFRs of S16. Although the scatter is still large, 0.37 dex, and manly due to the D4000-derived SFRs, the calibration is able to largely correct for the effects discussed above ans shown in Fig. \ref{app3} and \ref{app4}.

Fig. \ref{app5} shows the comparison of the Chang et al. (2015) SFR based on SED fitting results of MAGHPHYS code, from GALEX to WISE data. As already reported by S16, for a large percentage of galaxies, the Chang et al. (2015) catalog provides largely underestimated values of the SFR. Although such galaxies are in 60\% of the cases classified as star forming systems in the BPT diagram, according to the classification reported in the MPA-JHU catalog, their SFR is in disagreement with the H$\alpha$ based and IR based SFR estimates by several order of magnitudes. S16 discuss that this is likely due to the fact that the Magphys SED fitting results are mostly driven by the higher SNR 12 $\mu$m data-point than the low SNR 22 $\mu$m, leading to artificially extremely low SFR.

\begin{figure}
\includegraphics[width=0.9\columnwidth]{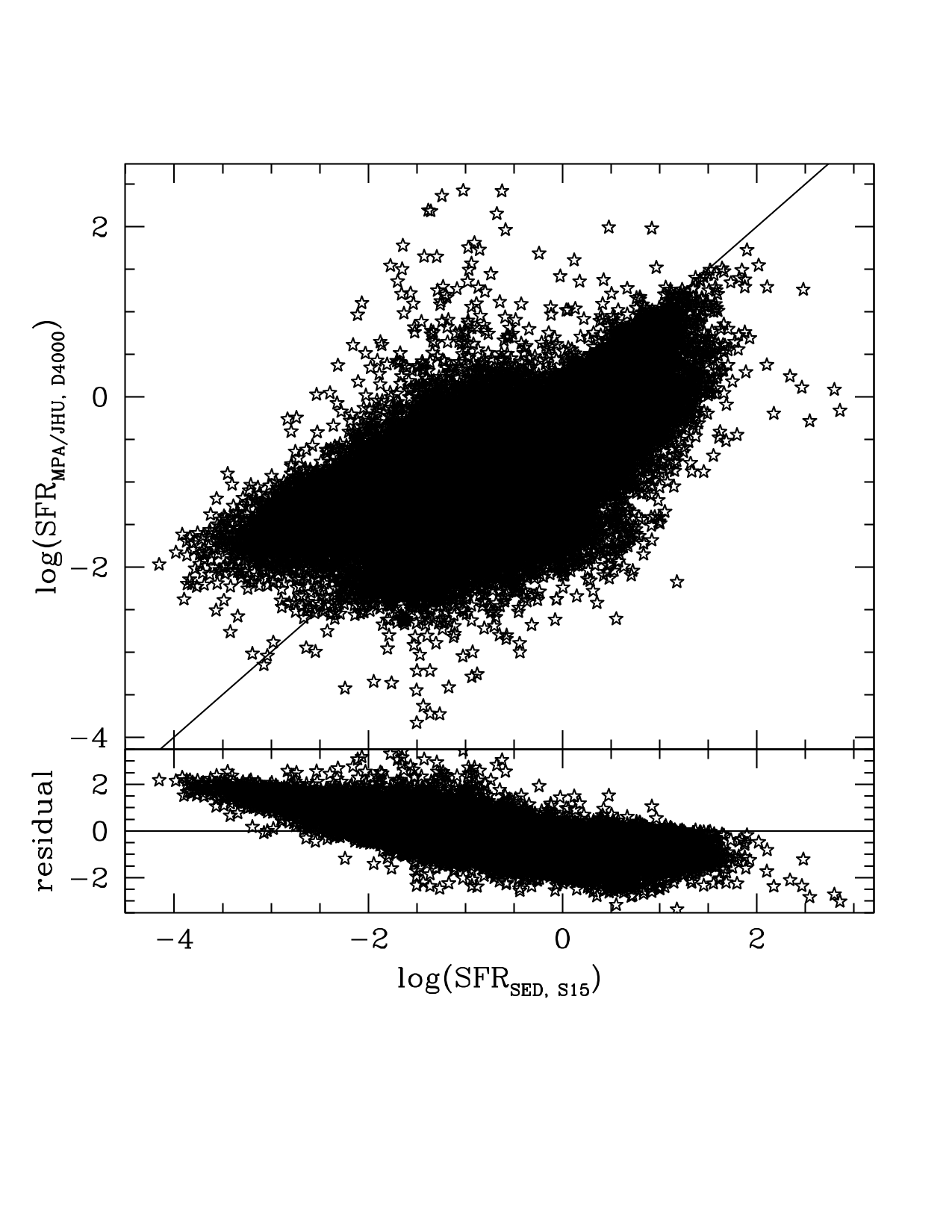}
\caption{Comparison of the SED based SFRs of Salim et al. (2016) and the D4000 based SFRs of the MPA-JHU catalog.}
\label{app4a}
\end{figure}

\begin{figure}
\includegraphics[width=0.9\columnwidth]{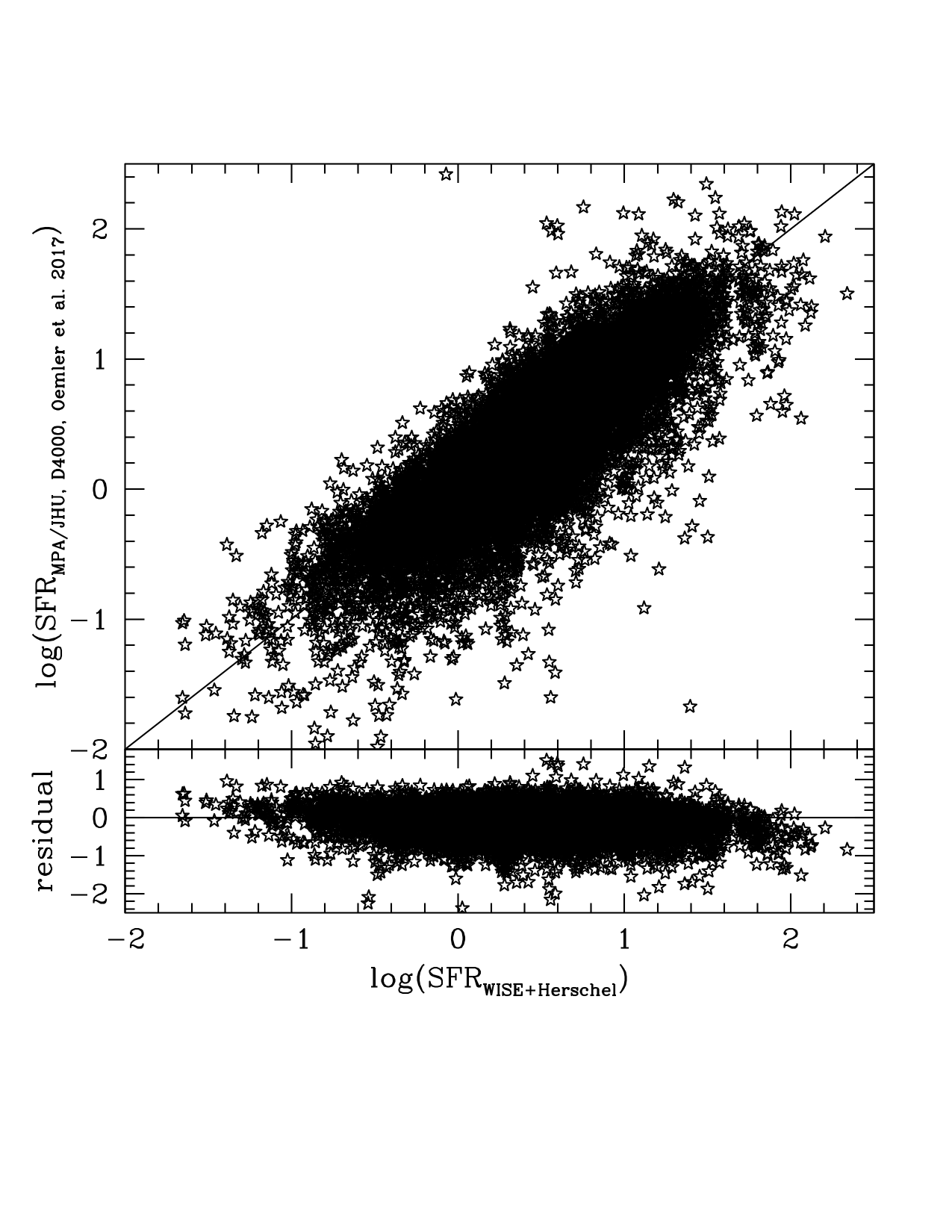}
\caption{Comparison of the SED based SFRs of Salim et al. (2016) and the D4000 based SFRs of the MPA-JHU catalog corrected according to the calibration of Oemler et al. (2017).}
\label{app4b}
\end{figure}

\begin{figure}
\includegraphics[width=0.9\columnwidth]{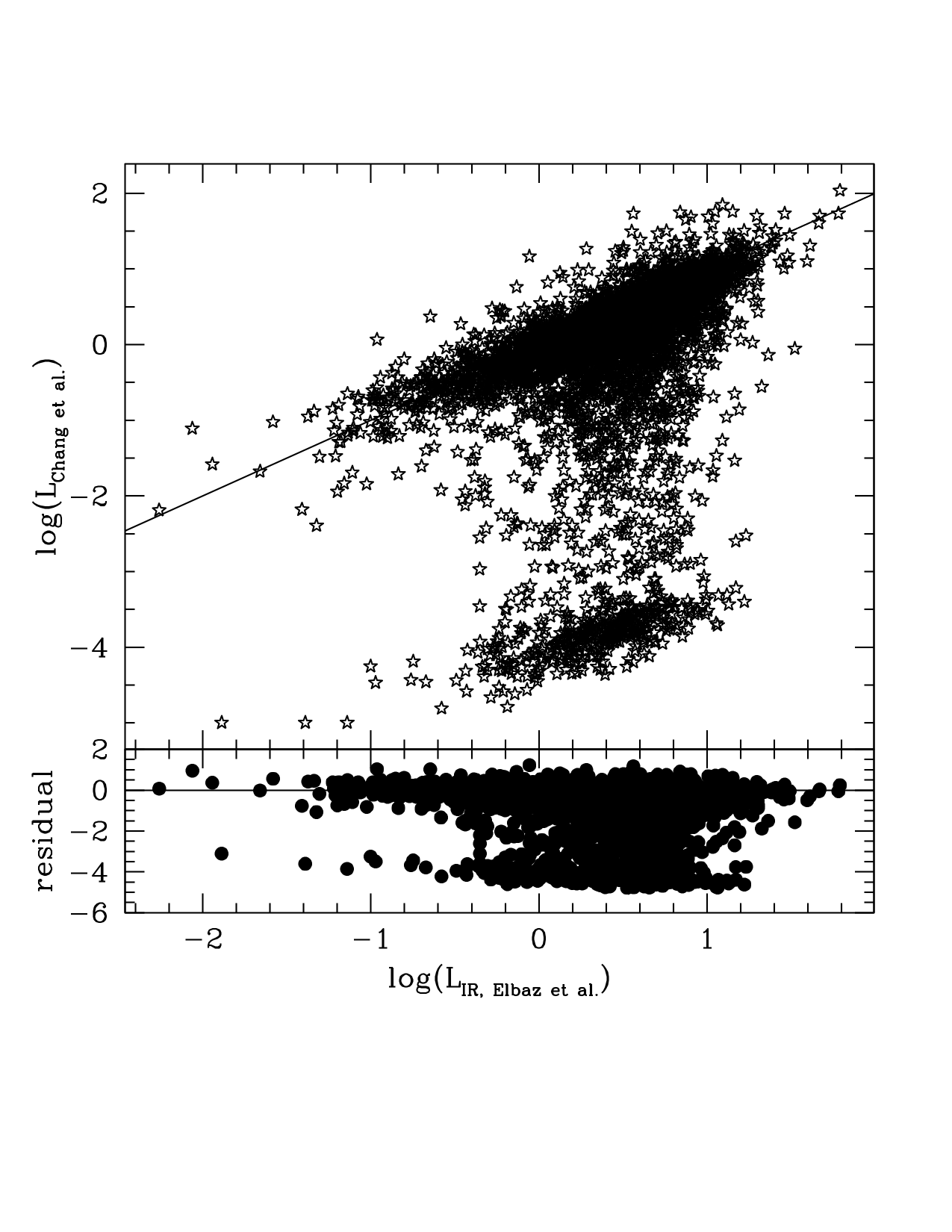}
\caption{Comparison of the SFR based on the far-infrared $L_{IR}$ derived in the low redshift sample with the Elbaz et al. (2011) MS and SB templates versus the SED fitting derived SFR of Chang et al. (2015) estimated with Magphys.}
\label{app5}
\end{figure}

\begin{figure*}
\includegraphics[width=0.9\columnwidth]{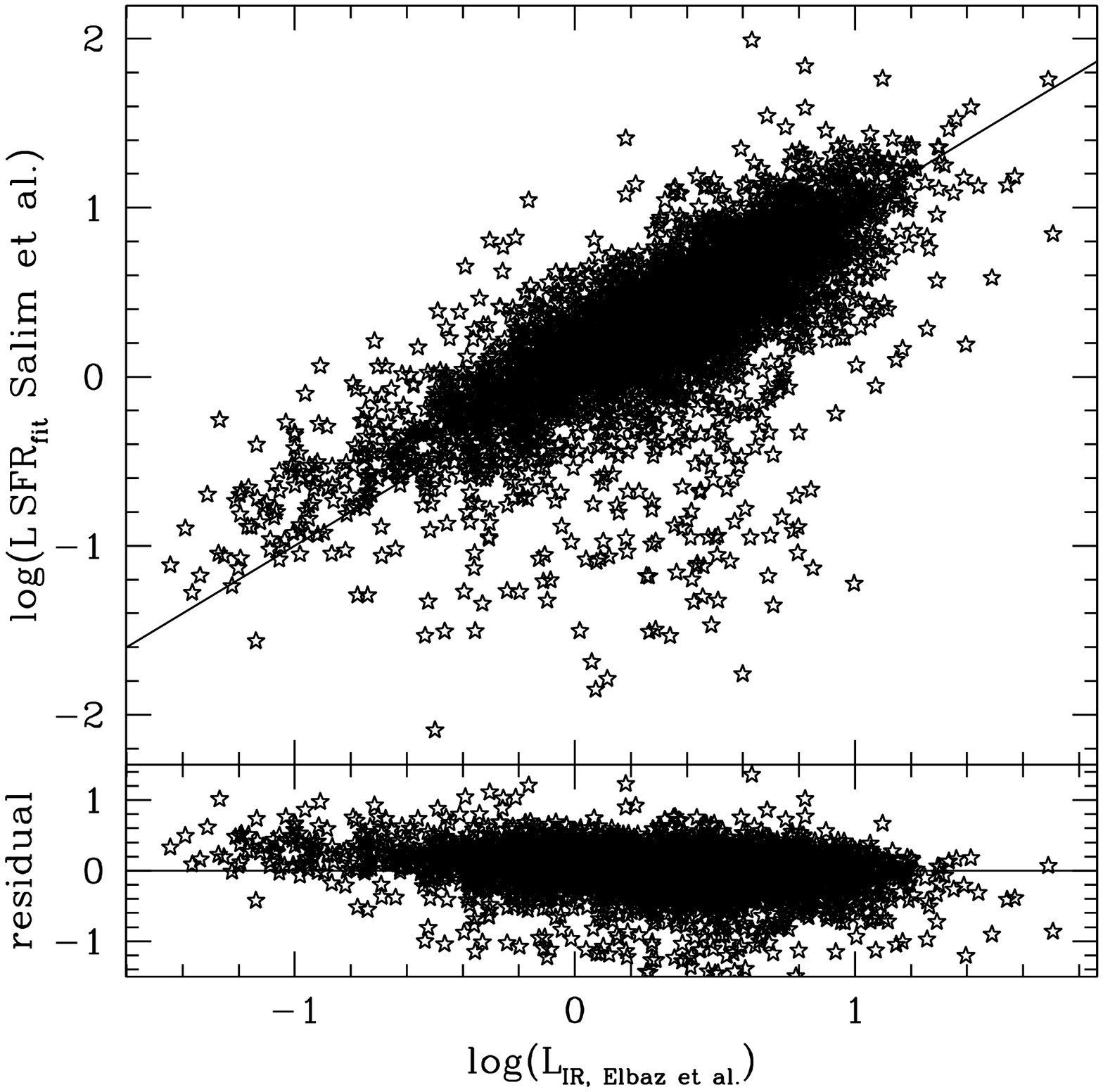}
\includegraphics[width=0.9\columnwidth]{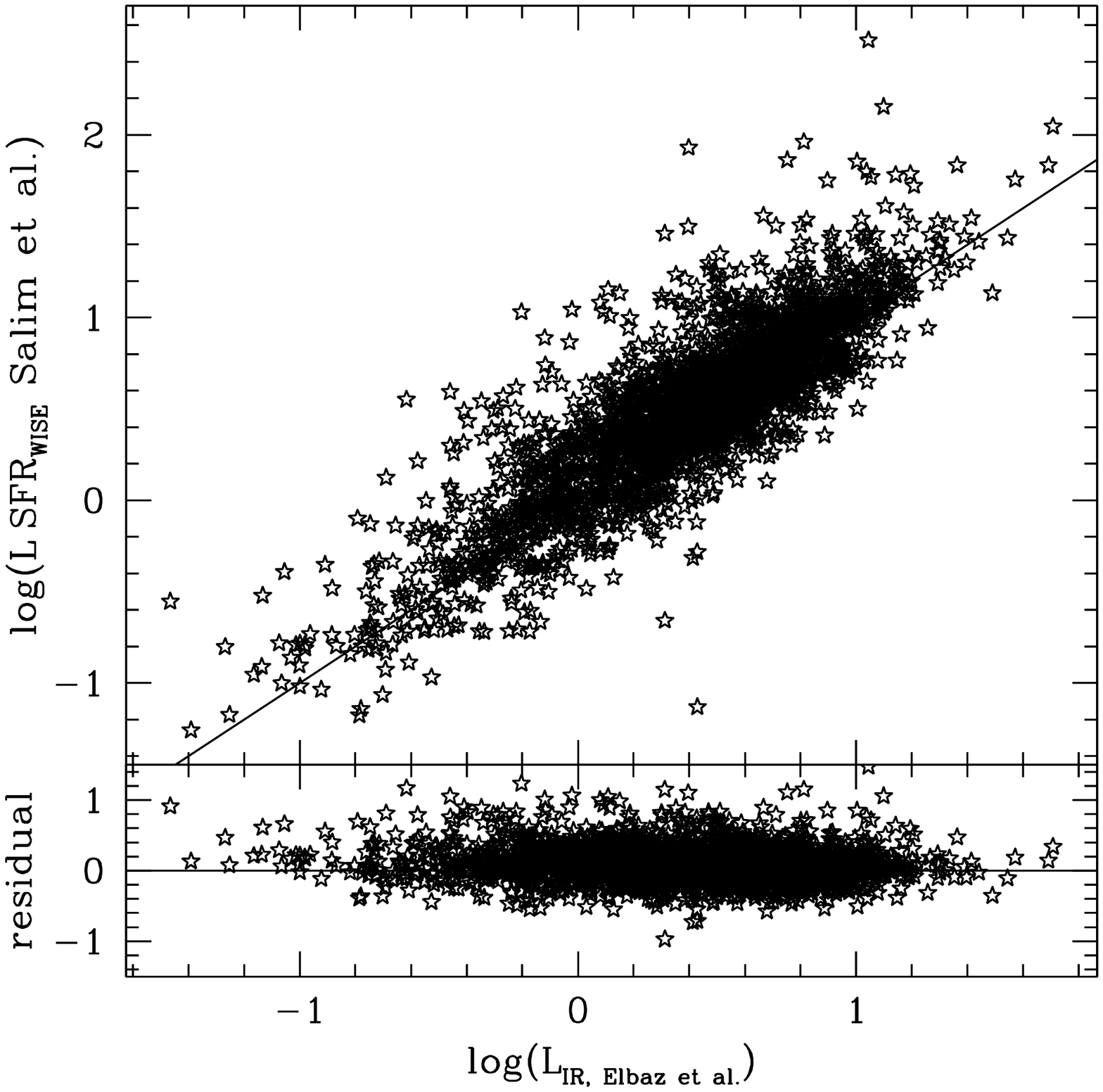}
\caption{{\it{Left panel}}: comparison of the SFR based on the far-infrared $L_{IR}$ derived in the low redshift sample with the Elbaz et al. (2011) MS and SB templates versus the SFR based on mid-infrared $L_{IR}$ derived from the WISE 22 $\mu$m data-point of S16. {\it{Right panel}}: comparison of the SFR based on the far-infrared $L_{IR}$ derived in the low redshift sample with SED fitting derived SFR of S16 estimated with Cigale.}
\label{app6}
\end{figure*}

Fig. \ref{app6} shows the comparison of reference SFR with the S16 catalog SFR estimates. S16 provide SFRs based on SED fitting with {\it{Cigale}} from GALEX UV to near-infrared data and an alternative estimate of the SFR based on the WISE 22 $\mu$m data only, when available. The right panel shows the comparison between the WISE based SFRs and the far-infrared derived SFRs. The agreement is as good as for the H$\alpha$ SFRs with a rms of 0.18 dex. We find a larger scatter between the SED fitting derived SFRs and the far-infrared based SFRs with a scatter of 0.46 dex. In addition we find that the SFRs based on SED fitting, mainly driven by the GALEX UV data, tend to be underestimated towards high value of SFR. This is more clearly visible in Fig. \ref{app7}, which shows the ratio between the SED fitting based SFRs and the far-infrared based SFRs versus the distance from the MS of RP15 up to stellar masses of $10^{10.5}$ $M_{\odot}$. The SFRs based on the SED fitting tend to be underestimated at larger distances from the MS location towards the upper envelope. This shows that SFRs based on UV and optical data only, despite the dust correction, are not able to capture the star formation activity of the dusty objects that tend to populate the upper envelope of the MS (Saintonge et al. 2011).

Similarly to Oemler et al. (2017), we also explore the dependence of the SFR indicators on the galaxy inclination. The SFRs of S16 based on WISE data do not show any dependence on the galaxy inclination. The H$\alpha$ and the D4000 based SFRs calibrated according to Oemler et al. (2017) include by construction the correction for the inclination dependence. We point out that this correction applies, in particular, to galaxies that in the original MPA-JHU catalog would lie in the lower envelope of the MS, as shown in Morselli et al. (2017). The UV based SFR derived through SED fitting do not show a clear trend with the inclination. Indeed, the obscuration due to the auto-extinction of the inclined disk is mixed with the effect shown in Fig. \ref{app7} for very dusty objects above the MS. Thus, a clean calibration is not easily applicable.

\begin{figure}
\includegraphics[width=0.9\columnwidth]{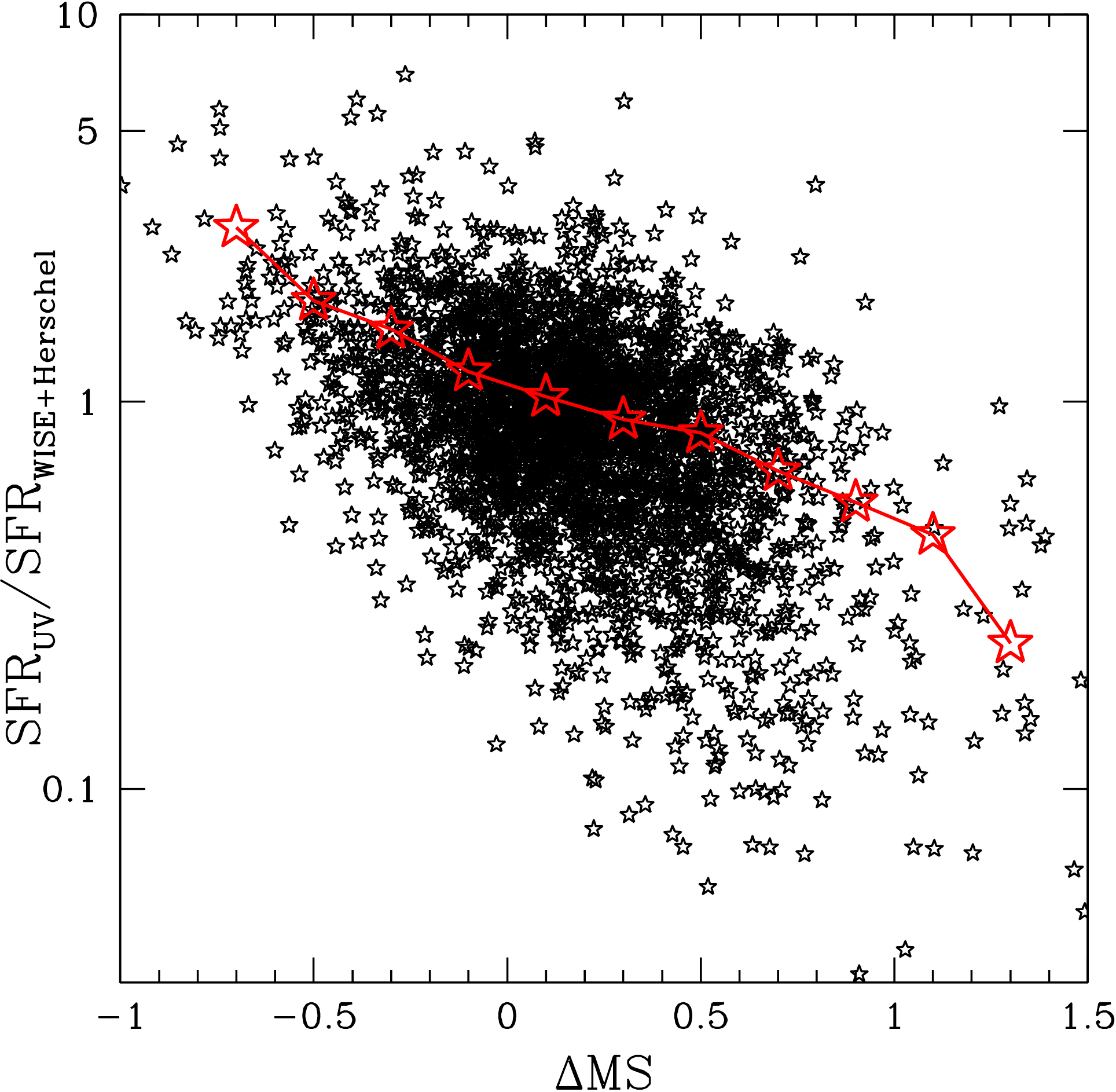}
\caption{Relation between the ratio $SFR_{SED fit}/SFR_{IR}$ of S16 SFR based on the SED fitting results  and the far-infrared based SFR based on WISE, PACS and SPIRE data versus the distance from the MS of RP15 (black points). The red stars indicate the median $SFR_{SED fit}/SFR_{IR}$}
\label{app7}
\end{figure}


\bsp	
\label{lastpage}
\end{document}